\pdfoutput=1
\documentclass[11pt,a4paper]{article}
\usepackage{jheppub,color,amsfonts,mathrsfs}
\usepackage[caption=false]{subfig}
\usepackage[titletoc,title]{appendix}

\newcommand{\beq}{\begin{eqnarray}}
\newcommand{\eeq}{\end{eqnarray}}
\newcommand{\non}{\nonumber\\}

\newcommand{\p}{\partial}

\newcommand{\tr}{\qopname\relax o{tr}}

\newcommand{\bpi}{\boldsymbol{\pi}}
\newcommand{\bsigma}{\boldsymbol{\sigma}}

\newcommand{\bx}{{\mathbf{x}}}
\newcommand{\bX}{{\mathbf{X}}}
\newcommand{\bn}{{\mathbf{n}}}
\newcommand{\bR}{{\mathbf{R}}}

\newcommand{\Og}{\qopname\relax o{O}}
\newcommand{\Lag}{\mathcal{L}}
\newcommand{\vol}{\qopname\relax o{vol}}
\newcommand{\SU}{\qopname\relax o{SU}}

\renewcommand{\i}{\mathrm{i}}
\renewcommand{\d}{\mathrm{d}}

\newcommand{\calF}{\mathcal{F}}
\newcommand{\calG}{\mathcal{G}}

\title{Dielectric Skyrmions} 

\author{Sven Bjarke Gudnason}
\affiliation{Institute of Contemporary Mathematics, School of
  Mathematics and Statistics, Henan University, Kaifeng, Henan 475004,
  P.~R.~China}
\emailAdd{gudnason(at)henu.edu.cn}

\abstract{
  We consider the dielectric Skyrme model proposed recently, with and
  without the addition of the standard pion mass term. Then we write
  down Bogomol'nyi-type energy bounds for both the massless and massive
  cases. We further show that, except for when taking the strict BPS
  limit, the Skyrmions are made of 3 orthogonal dipoles that can
  always be placed in their attractive channel and form bound states.
  Finally, we study the model numerically and discover that, long
  before realistic binding energies are reached, the Skyrmions become
  bound states of well-separated point-particle-like Skyrmions.
  By going sufficiently close to the BPS limit, we are able to obtain
  classical binding energies of realistic values compared with
  experiments. 
}

\begin{document}
\maketitle

\section{Introduction}

The Skyrme model \cite{Skyrme:1961vq,Skyrme:1962vh} is an effective
field theory with the symmetries of the strong interactions at low
energies, much like chiral perturbation theory
\cite{Weinberg:1978kz}. 
Two main differences between the approaches of the Skyrme model and
chiral perturbation theory lie in how the nucleon (baryon) is
implemented and that some fine tuning between some coefficients is
usually adopted in the Skyrme-type models.
The implementation of the nucleon in chiral perturbation theory is
simply done by means of a point-particle operator, as is standard in
fundamental quantum field theory, whereas in the Skyrme-model
approach, the nucleon is a topological soliton in the pion fields.
Clearly, taking the nucleon to be a point particle, is just an
approximation, but for many purposes a quite good one at sufficiently
low energies. The higher the energies are for the questions one is
asking, the worse it gets of course.
The other difference mentioned above, is that in chiral perturbation
theory, several operators at $p^4$ (fourth order in derivatives)
include four time derivatives, and in the Skyrme model it is customary
to take the coefficients of said operators such that the four time
derivatives cancel out, exactly.
This is a matter of practicality as it makes it possible to quantize
the spin and isospin zero modes by means of a standard Hamiltonian
analysis.

Serious attention was first given to the Skyrme model after Witten
pointed out that the baryon in large-$N$ QCD (Quantum Chromodynamics)
should be identified with the Skyrmion
\cite{Witten:1983tw,Witten:1983tx}. 
In particular, Adkins--Nappi--Witten provided the basis upon which many 
papers could continue the investigation of the Skyrme model as a model
of the nucleon \cite{Adkins:1983ya}, see the review
\cite{Zahed:1986qz}.
The first obstacle of studying nuclei beyond the single nucleon was to
obtain solutions to the full partial differential equations 
(PDEs), being the equations of motion (EOM) of the Skyrme model,
without spherical symmetry restrictions.
This was overcome by Battye and Sutcliffe by means of the rational map
approximation (RMA) and shell-like fullerenes of Skyrmions were
obtained \cite{Battye:1997qq}.
It was now clear that we were faced with a new problem: the binding
energies of the Skyrmions were far too large compared to those of
nuclei.
That is, typical binding energies in the Skyrme model are of the order
of 10\% per nucleon (Skyrmion), which should be compared to about 1\%
for real world nuclei. 
This binding-energy problem has been a theme in Skyrmion research for
the last 20+ years. 

The main tool for addressing the binding-energy problem is to write
down a topological energy bound or Bogomol'nyi bound for the static
energy of the model at hand \cite{Harland:2013rxa,Adam:2013tga}.
Such a bound exists for the standard (massless) Skyrme model and is
called the Skyrme--Faddeev bound \cite{Skyrme:1962vh,Faddeev:1976pg}
and reads $E\geq12\pi^2|B|$, 
where $B$ is the topological degree or baryon number. 
The energy bound, unfortunately, cannot be saturated by Skyrmions on
flat space, although it can be saturated by a Skyrmion on the
3-sphere, as shown by Manton and Ruback \cite{Manton:1986pz}. 
The Skyrmion with baryon number $B$ will be denoted as a $B$-Skyrmion 
throughout the paper.
The 1-Skyrmion has an energy that is about 23\% above the
Faddeev-Skyrme bound.
Now let us assume that the $B$-Skyrmion is bound.
This means that its energy divided by $B$ is smaller than 23\% over
the bound.
The $B$-Skyrmion with the largest possible binding energy would thus
correspond to the situation that it exactly saturates the bound.
This means that the closeness of the 1-Skyrmion to the energy bound
sets an upper limit on how large the binding energies can be for
multi-Skyrmions (i.e.~Skyrmions with $B>1$).

The above facts thus invited the community to a hunt for Skyrme-type
models that either have a saturable energy bound or have Skyrmion
solutions that can be quite close to the bound in some part of the
parameter space.
We shall use the expressions: energy bound, Bogomol'nyi bound and BPS
bound interchangeably throughout the paper. 
The models that achieve the above-stated goal can be classified into 3
different categories: (1) models that have an attainable BPS bound for
all topological charge sectors $B$; (2) models that have an attainable 
BPS bound only for the $B=1$ Skyrmion (with spherical symmetry); (3)
models that have an asymptotic series of terms, which when all added
up, hypothetically provide a model with an attainable BPS bound. 

In the first category of models attempting to solve the binding-energy
problem, there is only the BPS-Skyrme model
\cite{Adam:2010fg,Adam:2010ds} by
Adam--Sanchez-Guillen--Wereszczynski, which is a radical modification
of the Skyrme model. 
That is, the Skyrme model Lagrangian is replaced by a different
Lagrangian containing only a sixth-order derivative term, which is the
baryon charge density (current) squared, as well as a suitable
potential.
This model has Skyrmion solutions of all topological degrees
saturating the BPS bound.
Although it may seem a radical and perhaps almost contrived step to
replace the well-known terms of the chiral Lagrangian, this model has
been shown to be equivalent to the small 't Hooft coupling limit of
the Sakai-Sugimoto model \cite{Sakai:2004cn}, where the (holographic)
Skyrmions become large and fluid-like due to a dominating sextic
derivative term \cite{Bartolini:2017sxi}. 
It is, however, questionable that the known solutions with $B>1$ are
good candidates for near-BPS solutions, see
ref.~\cite{Speight:2014fqa} for the notion of restricted harmonicity
and ref.~\cite{Gudnason:2020tps} for near-BPS solutions in the
2-dimensional baby Skyrme model
\cite{Leese:1989gi,Piette:1994jt,Piette:1994ug}, which
is a toy model for the Skyrme model. 

In the second category, there is a Skyrme-type model made of the
Skyrme term and the pion mass term to the fourth power
\cite{Harland:2013rxa}, which was indeed proposed by Harland due to
the existence of a saturable energy bound that was found using the
H\"older inequality. 
Being in the second category, it has a Skyrmion solution that
saturates the energy bound, but only for $B=1$.
This has the imminent repercussion that all $B>1$ solutions are
unbound.
Nevertheless, small perturbations of the model can quickly remedy this
problem and provide a model of multi-Skyrmions with very low binding
energies \cite{Gillard:2015eia}, which was proposed by
Gillard--Harland--Speight. 
A modification of the Gillard--Harland--Speight model was proposed by
the author
\cite{Gudnason:2016mms,Gudnason:2016cdo,Gudnason:2016tiz,Gudnason:2018jia},
where instead of the potential $(1-n_0)^4$, the potential $(1-n_0)^2$
was used, which has the effect of letting the Skyrmions possess 
larger discrete symmetries with lower binding energies, compared to
the model of Gillard--Harland--Speight. 
Of course, for large enough potential coefficient, both models reduce
to the point-particle model of Skyrmions \cite{Gillard:2016esy}.
This limit, incidentally is physically equivalent to the large 't
Hooft coupling limit of holographic Skyrmions in the Sakai-Sugimoto
model \cite{Baldino:2017mqq}. 
Another model in the second category is made by introducing field
dependence to the coupling constants and is called the dielectric
Skyrme model \cite{Adam:2020iye} by Adam--Oles--Wereszczynski, which
is a special case of the Ferreira model \cite{Ferreira:2017bsr} where
the coupling constants are promoted from scalars to field-space
matrices (i.e.~the diagonal identity matrix with a scalar function
thus reduces to the latter model).
These models can be viewed as special cases in the framework put
forward in ref.~\cite{Adam:2013hza}. 
A very similar model has been proposed by Ferreira--Shnir
\cite{Ferreira:2017yzy} and Naya--Oles \cite{Naya:2020qvx}, where the
coupling constants are not functions of the fields, but of spatial
coordinates, which is in the spirit of the inclusion of impurities
\cite{Tong:2013iqa}. 
Models with field dependent coupling constants, have long been
considered, especially in the field of vortices, see
e.g.~refs.~\cite{Bazeia:2010wr,Bazeia:2012uc,Bazeia:2012ux} for a few first
examples. 

In the third category, there is the Sutcliffe model which is similar
in spirit to holographic QCD in that it derives from a Yang-Mills
action in five dimensions \cite{Sutcliffe:2010et,Sutcliffe:2011ig}. 
The main conceptual difference between the Sutcliffe model and a
holographic model like the Sakai-Sugimoto model \cite{Sakai:2004cn},
is that the latter has a scale -- the curvature of anti-de Sitter
space, which is  translated via the holographic dictionary to the QCD
scale (which in turn is defined as where the running gauge coupling of
QCD becomes nonperturbative).
Upon dimensional reduction from 5 dimensions to 4 dimensions, the
Yang-Mills theory can be written as the Skyrme model -- with
coefficients fixed by the ``background'' -- coupled to an infinite
tower of vector mesons, just like in the Sakai-Sugimoto model.
The missing gap in the Sutcliffe model is induced by truncating the
infinite tower of vector mesons to a finite number and re-calibrating 
the units, whereas in the Sakai-Sugimoto model, it is an intrinsic
quantity.
The beauty of the Sutcliffe model, is that it explains why the
instanton holonomy construction of Atiyah--Manton works so well.
That is, in the limit of including the entire tower of vector mesons,
the instanton holonomy description of the Skyrmion becomes exact.
Unfortunately, in this limit the theory also loses its intrinsic mass
scale, but it does become a BPS theory -- Yang-Mills theory in 5
dimensions.
In this sense, this model belongs to the third category; only when an
infinite tower of vector mesons is taken into account, the theory
becomes a BPS theory and hence has solutions that saturate the
Bogomol'nyi bound.

Apart from the BPS race, some noteworthy studies of other aspects of
physics, in particular of nuclear spectra, have been carried out.
First, however, the importance of the pion mass for Skyrmion solutions
with $B\gtrsim12$ was recognized, which ruled out the shell-like
fullerene structures as Skyrmion solutions for large atomic numbers
\cite{Battye:2004rw,Battye:2006tb,Houghton:2006ti}.
This led to the conclusion that smaller ``shells'' must be
preferred energetically, which turned out to be $B=4$ cubes happily
overlapping with the alpha particle model of nuclei
\cite{Battye:2006na}. 
Zeromode quantization of the classical Skyrmion solution, using the
rigid-body quantization put forward by Atkins--Nappi--Witten enjoyed
some success for light nuclei, but mostly so for even atomic numbers
(bosonic states) \cite{Battye:2009ad,Lau:2014baa}.
It turns out that in order to even get the ground state right for the
$B=7$ Skyrmion -- identified with lithium-7/beryllium-7 -- vibrational
modes have to be included in the low-energy spectrum
\cite{Halcrow:2015rvz}, thus proving the insufficiency of restricting
to zeromode quantization. 
Vibrational quantization was then tailored to specific nuclei with
quite some success, based judiciously on constructed toy models with
the right symmetries.
This was done for carbon-12 \cite{Rawlinson:2017rcq}, and oxygen-16 
\cite{Halcrow:2016spb,Halcrow:2019myn}. 
A further important effect was discovered, namely that it is not
enough to simply include both zeromodes and massive (vibrational)
modes of the Skyrmion in a quantization scheme, but mixing between them
leads to the crucial Coriolis effect \cite{Rawlinson:2019xsn}, which
improved the spectra for $B=4$ and $B=7$.
Finally, a systematic investigation of the low-energy vibrational
modes of Skyrmions with $B=1$ through $B=8$ was carried out
\cite{Gudnason:2018ysx}, laying the ground work for the development of
a better quantization scheme.

In this paper, we will study the classical solutions of another
variant of the Skyrme models that can be pushed close to a Bogomol'nyi
bound, namely the dielectric Skyrme model \cite{Adam:2020iye}
mentioned above. 
It is the model where the coupling constants, i.e.~the pion decay
constant, $F_\pi$, and the Skyrme coupling $e_{\rm Skyrme}$, are
promoted from constants (or running constants) to functions with field
dependence. 
It is easy to find the correct form of the product of the two coupling
functions, which unfortunately tends to zero in the vacuum in the BPS
limit.
It is unfortunate, because $e_{\rm Skyrme}=0$ makes the Skyrme term
ill-defined and as well known this term (if not other
higher-derivative terms are added to the Lagrangian) is a necessity
for stabilizing the size of the Skyrmion due to Derrick's theorem
\cite{Derrick:1964ww}.
As for $F_\pi=0$ in the vacuum, this makes the pions nonpropagating in
the vacuum and hence physically not acceptable.
The damage is not too severe, since we do not want a BPS model anyway,
we just want to move sufficiently close to such a limit, and hence we
will consider the proposal by Adam--Oles--Wereszczynski to freeze the
pion decay constant during its descent \cite{Adam:2020iye}, such that
it attains a finite value in the vacuum -- thus allowing pions to
propagate. 
Of course, there is some arbitrariness in how the freezing is done,
but we will just stick to the simplest possibility in this paper.

At this point we have summarized, in quite some detail, some of the
efforts that have been made for pushing the Skyrme program towards a
working framework for low-energy nuclear physics.
It would be illuminating at this point to pause and consider what we
are trying to do in more abstract terms.
We assume that the standard model of particle physics describes all
known baryonic matter and its interactions.
The most crucial ingredient is QCD, which is a Yang-Mills theory with
gauge group SU(3).
The particles charged under SU(3) of the strong interactions, are
the quarks and the gluons (gauge fields of SU(3)).
QCD has the unfortunate feature of possessing asymptotic freedom,
which means that perturbation theory can be utilized only at large
energies (say above 10 GeV or higher) and is useless at the QCD scale
(about 250 MeV), below which the spectra of nuclei have to be
extracted.
The approach we take is called effective field theory, and in
principle it is clear what has to be done.
We should identify the scale at which we are interested in asking our
questions and which particles are the low-energy degrees of freedom
and further what symmetries they possess.
The answer is also clear.
Energies less than 10-20 MeV would be sufficient, and the symmetry
and particle content is $\SU(2)_L\times\SU(2)_R$ chiral symmetry and
pions, respectively.
Now the recipe is simple, write down all possible symmetry-preserving
operators in the Lagrangian up a high enough order in mass-dimension.
The difficulty is that the pions were not in the QCD Lagrangian to
begin with, but are composite particles made of two quarks in a bound
state -- bound by gluons.
Therefore, we do not know how to determine the coefficients of the
operators in the chiral Lagrangian from a theoretical point of view,
they must be determined experimentally -- hence with experimental
uncertainties.\footnote{There are ways to determine almost all
  coefficients (low-energy constants): Either by assuming hidden local
  symmetry \cite{Harada:2003jx} or by assuming a specific holographic
  background, like in the Sakai-Sugimoto model \cite{Sakai:2004cn} or
  the Sutcliffe model \cite{Sutcliffe:2010et}. } 
The other problem arises due to the nature of using a soliton to
describe the nucleon.
That is, the soliton is an extended field configuration that even in
the ground state has nonvanishing field derivatives.
This makes it difficult to justify any order (in mass dimension or in
derivatives) to which we may truncate the effective theory, and hence
motivates a theory with an infinite number of terms (for a guess see
refs.~\cite{Marleau:1989fh,Marleau:1991jk,Marleau:2000ia}). 
It is, nevertheless, plausible that resummations of terms are possible 
in certain sectors and that due to such resummations, the infinite
tower of operators can -- at least approximately -- be written as a
sum of geometric series multiplying fundamental operators of the
Lagrangian.
Such a wild-eyed philosophy is what we will offer as a justification
for allowing the coupling constants to be functions of the fields of
the model.

The paper is organized as follows.
The dielectric Skyrme model is reviewed in sec.~\ref{sec:model} and
its Bogomol'nyi bound in sec.~\ref{sec:Bogomolnyi}. The near-BPS
version of the model is described in sec.~\ref{sec:nearBPS}, the
inclusion of the pion mass term in sec.~\ref{sec:pionmass} and the
modification of the energy bound due to the pion mass term in
sec.~\ref{sec:energybound}. Then the equations of motion are studied
in sec.~\ref{sec:eom} and linearized to reveal the single Skyrmion as
3 orthogonal dipoles, which can attract another Skyrmion in the
attractive channel in the entire parameter space of the near-BPS
model, see sec.~\ref{sec:forces}. The numerical results are presented
in sec.~\ref{sec:num} and finally the paper is concluded in
sec.~\ref{sec:discussion} with a discussion.

\section{The dielectric Skyrme model}\label{sec:model}

We consider the model \cite{Adam:2020iye}
\beq
\Lag = \frac{f^2}{2}\tr(L_\mu L^\mu)
+\frac{1}{16e^2} \tr\left([L_\mu,L_\nu][L^\mu,L^\nu]\right),
\label{eq:L}
\eeq
with $f=f(\tr U/2)$ and $e=e(\tr U/2)$ being dimensionless
functions, we have defined the left-invariant chiral current 
\beq
L_\mu \equiv U^\dag \p_\mu U,
\eeq
the field $U$ takes value in $\SU(2)$ and is related to an
$\Og(4)$-vector $\bn=(n_0,n_1,n_2,n_3)$ by
\beq
U = n_0\mathbf{1}_2 + \i\tau^a n_a, \qquad a=1,2,3,
\eeq
where $(n_1,n_2,n_3)$ are known in physics as pions and $\tau^a$ are
the Pauli matrices.
The spacetime indices $\mu,\nu$ run over $0,1,2,3$ and we use the flat
Minkowski metric of the mostly-positive signature.
The model as written in eq.~\eqref{eq:L} is already in Skyrme units,
where energy is measured in units of $F_\pi/(4e_{\rm Skyrme})$
and lengths are measured in units of $2/(F_{\pi}e_{\rm Skyrme})$, with
$F_\pi$ the pion decay constant and $e_{\rm Skyrme}$ the Skyrme
coupling \cite{Manton:2004}.  

The functions $f$ and $e$ explicitly break the symmetry from
$\SU(2)\times\SU(2)$ down to $\SU(2)$ (diagonal or vectorial).
The theory is ungauged and therefore the requirement of finite energy
configurations amounts to enforcing $\p_\mu U=0$ at spatial infinity,
which in turn effectively point compactifies 3-space from
$\mathbb{R}^3$ to $\mathbb{R}^3\cup\{\infty\}\simeq S^3$.
$\SU(2)$ viewed as a manifold is isomorphic to $S^3$ and hence the
theory is characterized by the topological degree
\beq
\pi_3(S^3) = \mathbb{Z} \ni B,
\eeq
where $B$ is the degree of the mapping $U$ and is identified with the
number of baryons.
The topological degree can be directly calculated as
\beq
B = -\frac{1}{24\pi^2}\int_{\mathbb{R}^3} \epsilon_{ijk}
\tr(L_iL_jL_k) \;\d^3x,
\label{eq:B}
\eeq
where $i,j,k$ are spatial indices and thus run only over
$1,2,3$ and we adopt the convention $\epsilon_{123}=+1$.

\subsection{Bogomol'nyi bound}\label{sec:Bogomolnyi}

The static energy can be written as\footnote{The Bogomol'nyi bound
  written in the form below is not so common, but has appeared
  several times in the literature, see
  refs.~\cite{Manton:1986pz,Ferreira:2013bia}. }
\begin{align}
E &= \int_{\mathbb{R}^3} \tr\left[
  -\frac{f^2}{2}L_iL_i
  -\frac{1}{16e^2}[L_i,L_j]^2
  \right]\d^3x\non
&= \int_{\mathbb{R}^3}\tr\left[
  -\frac12\left(fL_i \mp \frac{1}{2e}\epsilon_{ijk}L_jL_k\right)^2
  \mp \frac{f}{2e}\epsilon_{ijk}L_iL_jL_k
  \right]\d^3x,
\label{eq:Estatic}
\end{align}
where we have performed a Bogomol'nyi completion in the second line.
The BPS equations are thus
\beq
L_i = \pm \frac{1}{2f e}\epsilon_{ijk}L_jL_k,
\label{eq:BPS}
\eeq
yielding (anti-)Skyrmions for the upper (lower) sign. 
The fact that the cross term (last term) in the last line of the static
energy \eqref{eq:Estatic} is proportional to the topological degree --
even for $f$ and $e$ generalized to functions of $\tfrac12\tr U$ -- is due to
the fact that they are target space functions.
This can be seen as follows.
Rewriting the expression for the topological degree \eqref{eq:B}, we
have
\begin{align}
  B &= -\frac{1}{24\pi^2}\int_{\mathbb{R}^3} \epsilon_{ijk}
  \tr(U^\dag\p_i U U^\dag\p_j U U^\dag\p_k U) \; \d^3x \non
  &= \frac{1}{12\pi^2}\int_{\mathbb{R}^3}
  \epsilon_{A B C D}\epsilon_{ijk} \,
  n_A \p_i n_B\p_j n_C\p_k n_D \; \d^3x\non
  &= \int_{\mathbb{R}^3} \bn^*\vol_{S^3}
  = B \int_{S^3} \vol_{S^3},
\end{align}
where the indices $A,B,C,D$ run over $0,1,2,3$, we use the convention
$\epsilon_{0123}=+1$ and the normalized volume form on $S^3$ is
denoted by $\vol_{S^3}$.
The topological degree is simply the pullback of the volume form on
$S^3$ to $\mathbb{R}^3$ by the field (map) $\bn$ (or equivalently
$U$), denoted by $\bn^*$.

The integral of the cross term in the static energy \eqref{eq:Estatic} 
after Bogomol'nyi completion, can thus be written as
\begin{align}
  E_{\rm BPS} &= \pm12\pi^2\int_{\mathbb{R}^3} \frac{f(n_0)}{e(n_0)} \bn^*\vol_{S^3} \non
  &= 12\pi^2|B| \int_{S^3} \frac{f(n_0)}{e(n_0)} \vol_{S^3},
  \label{eq:E_BPS}
\end{align}
which can be interpreted as the target-space average of the function
$f/e$ on the 3-sphere \cite{Adam:2020iye}.

The BPS equation \eqref{eq:BPS} can be solved only for $B=1$
\cite{Adam:2020iye}, for which we can employ a hedgehog Ansatz
\beq
U = \cos\xi(r)\mathbf{1}_2 + \i\tau^a\hat{x}^a\sin\xi(r), \qquad a=1,2,3,
\label{eq:Uhedgehog}
\eeq
where we have defined $\hat{x}^a=x^a/r$ with $r=\sqrt{x^ax^a}$.
The left-invariant thus reads
\beq
L_i = \i\tau^i \frac{\sin 2\xi}{2r}
+\i\hat{x}^i\hat{x}^a\tau^a\left(\xi' - \frac{\sin2\xi}{2r}\right)
-\i\epsilon^{iab}\hat{x}^a\tau^b\frac{\sin^2\xi}{r},
\eeq
which is the left-hand side of the BPS equation \eqref{eq:BPS},
whereas the right-hand side, being the commutator of $L_j$ and $L_k$,
reads 
\begin{equation}
\pm\frac{1}{f e}
\left(-\i\tau^i\frac{\sin2\xi}{2r}\xi'
+\i\hat{x}^i\hat{x}^a\tau^a\left[\frac{\sin2\xi}{2r}\xi'
  - \frac{\sin^2\xi}{r^2}\right]
+\i\epsilon^{iab}\hat{x}^a\tau^b\frac{\sin^2(\xi)\xi'}{r}
\right),
\end{equation}
which yields
\beq
\xi' = \mp f e, \qquad
f e \xi' = \mp\frac{\sin^2\xi}{r^2},
\label{eq:B=1BPS}
\eeq
where the former equation is obtained from all three tensor structures
and the latter comes from the second tensor structure.
Using the former equation in the latter, we thus get
\beq
(\xi')^2 = \frac{\sin^2\xi}{r^2}.
\eeq
Taking the square root and choosing the negative sign and choosing the
upper sign in eq.~\eqref{eq:B=1BPS}, we have
\beq
\xi' = - f e, \qquad
\xi' = -\frac{\sin\xi}{r},
\eeq
whose common solution with the appropriate boundary conditions
$\xi(0)=\pi$ and $\xi(\infty)=0$ reads
\beq
\xi = 2\arctan\frac{r_0}{r}, \qquad
f e = \frac{1}{r_0}(1 - \cos\xi).
\eeq
This is exactly the $B=1$ BPS solution found in
ref.~\cite{Adam:2020iye}. 
Since the vacuum of the theory lies at $\xi=0$ or equivalently
$U=\mathbf{1}_2$, the product $f e$ vanishes in the vacuum in the
BPS limit.
In terms of $\bn$, we have
\beq
f(n_0) e(n_0) = \frac{1}{r_0}(1 - n_0).
\eeq

Obviously, we cannot allow $e(n_0)$ to tend to zero, as that will
make the Skyrme term ill defined in the vacuum.
We would also like $f(n_0)$ to take a nonvanishing value in the vacuum
so as to conform with the physics of perturbative pions, far from the
cores of nuclei.
This leads us to the consideration of a near-BPS model.

\subsection{Near-BPS dielectric Skyrme model}\label{sec:nearBPS}

The near-BPS version we will study in this paper is based on
freezing $f(n_0)$ at a finite value before $n_0$ reaches the vacuum
value $n_0=n_0^{\rm vac}=1$ \cite{Adam:2020iye}.
This will also allow us to introduce the normal pion mass term as a
secondary (additional) BPS breaking term, without facing the problem
of infinitely heavy pions.

The \emph{massless} near-BPS version of the model in $\bn$ coordinates reads
\beq
\Lag = - f^2(n_0)\p_\mu\bn\cdot\p^\mu\bn
- \frac12(\p_\mu\bn\cdot\p^\mu\bn)^2
+ \frac12(\p_\mu\bn\cdot\p_\nu\bn)(\p^\mu\bn\cdot\p^\nu\bn),
\eeq
where we have fixed the coupling functions of the model as
\beq
f(n_0) =
\begin{cases}
  \frac{1 - n_0 + \beta}{1 - n_\star + \beta}, & n_0 < n_\star,\\
  1, & n_0 \geq n_\star,
\end{cases}\qquad
e(n_0) = 1,
\eeq
with $\beta>0$ a positive constant.

This model has the nice feature that we can interpolate between the
standard (massless) Skyrme model, i.e.~for $n_\star=-1$, and the BPS
model, i.e.~for $n_\star=1$.
In the standard Skyrme model, the Skyrmions are too strongly bound,
whereas in the BPS limit there are only $B=1$ solutions that saturate
the BPS bound, which in turn implies that all $B>1$ Skyrmions are
unstable.  
We can therefore interpolate between a model that is too strongly
bound and a model that is unbound.
Somewhere in between $n_\star\in(-1,1)$ there may be realistic values
for the binding energies of the multi-Skyrmions.
What properties they have is what we want to study in this paper. 

We can now exclude the possibility of going strictly to the BPS limit
and therefore allow $n_\star\in[-1,1)$ only and hence
\beq
f(n_0) =
\begin{cases}
  \frac{1 - n_0}{1 - n_\star}, & n_0 < n_\star,\\
  1, & n_0 \geq n_\star,
\end{cases}\qquad
e(n_0) = 1,
\label{eq:fe_funs}
\eeq
where we have set $\beta=0$.
The normalization by $(1-n_\star)$ is practical for making sure the
Skyrmions do not grow to unreasonable sizes (which is impractical for
numerical calculations). 

We can now calculate the Bogomol'nyi bound \eqref{eq:E_BPS} for the
specific functions \eqref{eq:fe_funs}:
\begin{align}
  E_{\rm BPS} &= 12\pi^2|B|\left(
  \int_{S^3|n_0<n_\star}\frac{1-n_0}{1-n_\star}\vol_{S^3}
  +\int_{S^3|n_0\geq n_\star}\vol_{S^3}
  \right)\non
  &=12\pi^2|B|\left(
  \frac{2}{\pi(1-n_\star)}\int_{\arccos(n_\star)}^\pi2\sin^2\left(\frac{\xi}{2}\right)\sin^2\xi\;\d\xi
  +\frac{2}{\pi}\int_0^{\arccos(n_\star)}\sin^2\xi\;\d\xi
  \right)\non
  &=12\pi^2|B|\calF(n_\star),
  \label{eq:Bogomolny_nearBPS_massless}
\end{align}
where we have defined the function
\beq
\calF(n_\star) \equiv
\frac{(2+n_\star^2)\sqrt{1-n_\star^2}+3(\pi-n_\star\arccos(n_\star))}{3\pi(1-n_\star)}.
\eeq
The above defined function obeys $\calF(-1)=1$, whereas in the
limit $\lim_{n_\star\to 1^-}\calF(n_\star)\to\infty$, it
diverges.
This will not pose a problem, as we will not attempt at reaching the
BPS limit.
\begin{figure}[!htp]
  \begin{center}
    \includegraphics[width=0.5\linewidth]{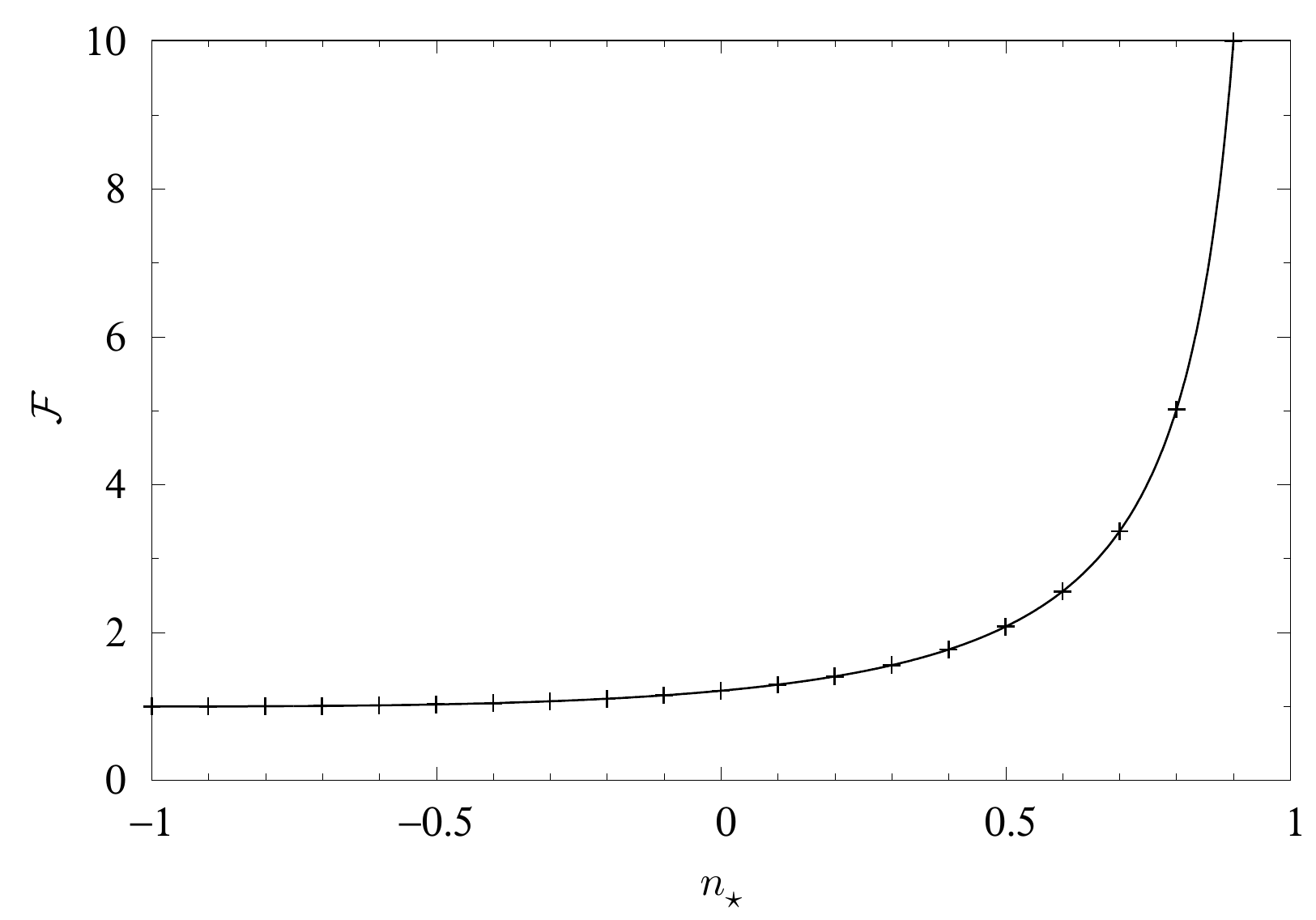}
    \caption{The functional behavior $\calF(n_\star)$ of the
      Bogomol'nyi bound \eqref{eq:Bogomolny_nearBPS_massless} in the
      massless case. The points (pluses) for every
      $n_\star\in\mathbb{Z}/10$, correspond to numerical solutions in
      later sections.
    }
    \label{fig:calF}
  \end{center}
\end{figure}
The function $\calF(n_\star)$ is shown in fig.~\ref{fig:calF}.

It will prove convenient to define an energy functional, that is
divided by the Bogomol'nyi bound
\beq
\epsilon_B \equiv \frac{E}{12\pi^2|B|\calF(n_\star)},
\label{eq:epsilon_B}
\eeq
which holds for the \emph{massless} near-BPS dielectric Skyrme model.

\subsection{Including the pion mass term}\label{sec:pionmass}

Although the near-BPS model discussed in the previous section, has the
aesthetically nice feature that we can interpolate between the
standard \emph{massless} Skyrme model and a BPS theory, we do want to
break the BPS-ness.
Hence introducing the standard pion mass term is not a problem.
However, this means that we interpolate between the standard
\emph{massive} Skyrme model and another near BPS-model, so it is no
longer guaranteed that we will reach low binding energies of
multi-Skyrmions.
Nevertheless, it is a straightforward numerical exploration and we
will thus add to the near-BPS dielectric model the mass term
\beq
V = m^2\tr(\mathbf{1}_2 - U)
= 2m^2(1 - n_0),
\label{eq:V}
\eeq
which is the standard normalization and the pion mass parameter
is often taken to be $m=1$ in Skyrme units \cite{Battye:2006na}.
Notice that because of our normalization of $f(n_0)$ such that
$f(n_0)=1$ in the pion vacuum, the pion mass term is correctly
normalized in this massive version of the near-BPS model. 

\subsubsection{Energy bound}\label{sec:energybound}

Including the pion mass term alters the bound on the energy.
In order to derive the bound, we first use the result of Harland
\cite{Harland:2013rxa}, for a submodel
\begin{align}
  E^{04} = \int_{\mathbb{R}^3}\left[
    -\frac{1}{16}\tr\big([L_i,L_j]^2\big) + V\right],
  \label{eq:E04}
\end{align}
which reads \cite{Harland:2013rxa}
\begin{align}
  E^{04} \geq 8\pi^2|B|\langle V^{1/4}\rangle, \qquad
  \langle V^{1/4}\rangle\equiv \int_{S^3} V^{1/4} \vol_{S^3},
  \label{eq:Harland_bound}
\end{align}
where we have set $e=1$.
Writing now the energy as
\begin{align}
  E = \int_{\mathbb{R}^3}\left[ - \frac{f^2}{2}\tr(L_iL_i)
    -\frac{1-\alpha+\alpha}{16}\tr\big([L_i,L_j]^2\big) + V\right],
\end{align}
and following refs.~\cite{Harland:2013rxa,Adam:2013tga}, we get a
combined bound
\begin{align}
  E \geq 12\pi^2|B|\calG, \qquad
  \calG \equiv \langle f\rangle\left(\sqrt{\alpha}
  +\frac{2}{3\sqrt{a}}(1-\alpha)^{3/4}\right), \qquad
  a \equiv \frac{\langle f\rangle^2}{\langle V^{1/4}\rangle^2}>0,
  \label{eq:combined_bound}
\end{align}
which is a combined bound of eq.~\eqref{eq:E_BPS} for the massless
dielectric Skyrme model and of eq.~\eqref{eq:Harland_bound} for the
submodel \eqref{eq:E04}.
The parameter $\alpha\in[0,1]$ is a weight of how much of the Skyrme
term is used in the normal Skyrme-Faddeev bound and how much is used
in the bound for the submodel \eqref{eq:E04}. 

The bound should be maximized by varying $\alpha$, which yields
\cite{Adam:2013tga} 
\beq
\widehat{\alpha} = \frac{a^2}{2}\left(\sqrt{1+\frac{4}{a^2}}-1\right).
\label{eq:widehat_alpha}
\eeq
Keeping $\langle f\rangle$ fixed and sending $m\to 0$ in $V$ of
eq.~\eqref{eq:V} yields the limit
\beq
\lim_{a\to\infty}\widehat\alpha = 1.
\eeq
Therefore, the combined bound \eqref{eq:combined_bound} correctly
reduces to the bound \eqref{eq:E_BPS} in the limit of $m\to0$ (because
$\widehat\alpha\to1$ and $a^{-1/2}\to0$).
Contrarily, if we keep $\langle f\rangle$ fixed and send $m\to\infty$,
$a$ goes to zero for which we have
\beq
\lim_{a\to0}\widehat\alpha = 0. 
\eeq
The bound \eqref{eq:combined_bound} still diverges, but since
$\widehat\alpha\to0$ only the part from the bound
\eqref{eq:Harland_bound} remains.

\begin{figure}[!htp]
  \begin{center}
    \includegraphics[width=0.5\linewidth]{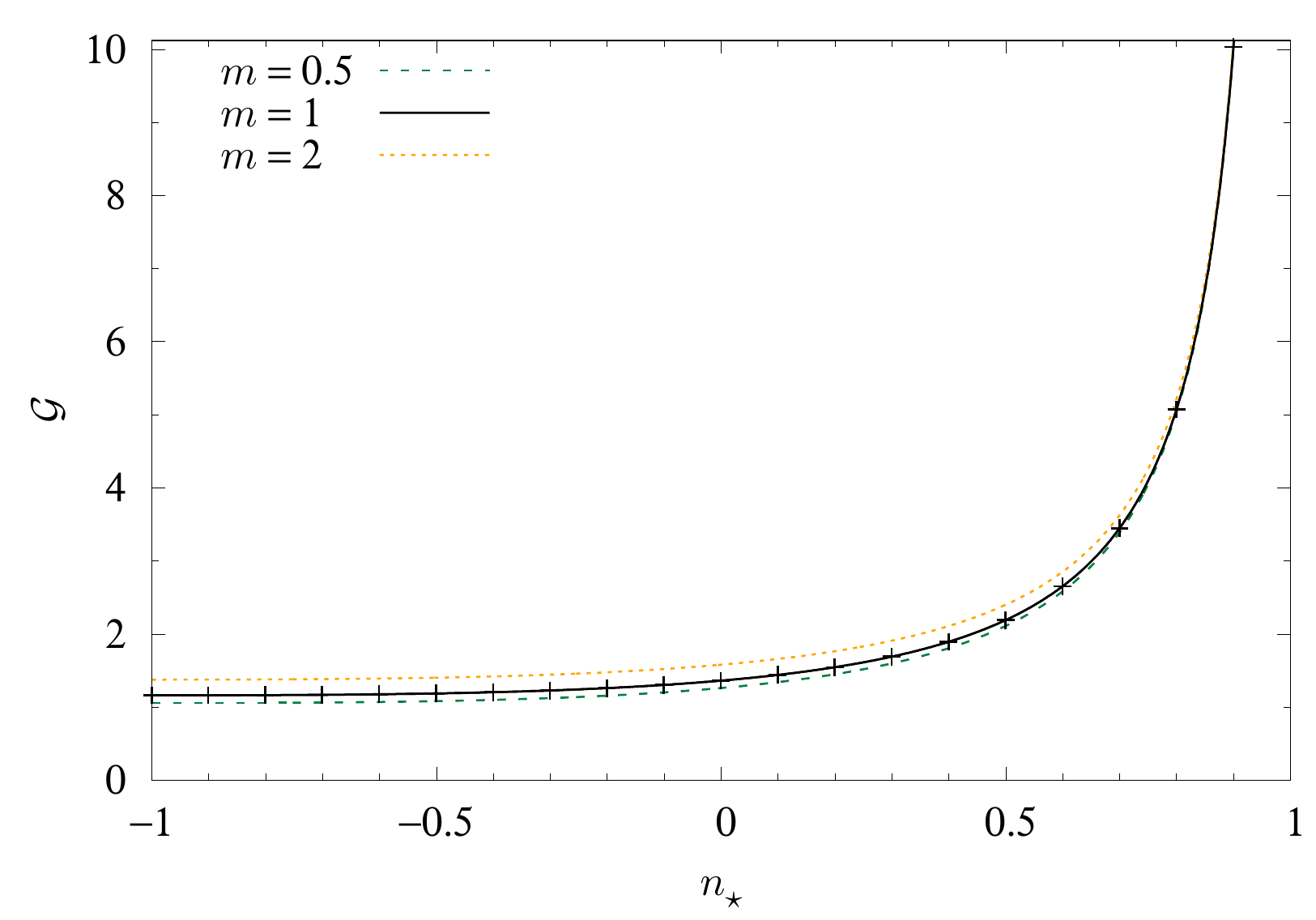}
    \caption{The functional behavior $\calG(n_\star,m)$ of the
      energy bound \eqref{eq:combined_bound_eval} in the
      massive case for various values of $m=0.5,1,2$. The points
      (pluses) for every $n_\star\in\mathbb{Z}/10$, correspond to
      numerical solutions in later sections (for $m=1$).
    }
    \label{fig:calG}
  \end{center}
\end{figure}

Evaluating now the target space integrals for the function $f$ of
eq.~\eqref{eq:fe_funs} and the potential $V$ of eq.~\eqref{eq:V}, we
get
\begin{align}
  E &\geq 12\pi^2|B|\calG(n_\star,m), \non
  \calG(n_\star,m) &=
  \calF(n_\star)\left(\sqrt{\widehat\alpha}
  +\frac{2}{3\sqrt{a}}(1-\widehat\alpha)^{3/4}\right), \non
  a &= \frac{225\pi^3\calF^2(n_\star)}{4096m\Gamma^4\left(\tfrac34\right)}
  \simeq 0.755 \frac{\calF^2(n_\star)}{m},
  \label{eq:combined_bound_eval}
\end{align}
where $\Gamma$ is Euler's gamma function and $\widehat\alpha$ is given
by eq.~\eqref{eq:widehat_alpha}. 
The function $\calG(n_\star,m)$ is shown in fig.~\ref{fig:calG}.

It will again be convenient to define an energy functional, that is
divided by the energy bound
\beq
\epsilon_B \equiv \frac{E}{12\pi^2|B|\calG(n_\star,m)},
\label{eq:epsilon_B_m1}
\eeq
which holds for the \emph{massive} near-BPS dielectric Skyrme model.

\subsection{Equations of motion}\label{sec:eom}

The full equations of motion, which are partial differential equations
(PDEs), read
\begin{align}
f^2(n_0)\p^2n_a
+2f(n_0)f'(n_0)\p_\mu n_0\p^\mu n_a
+(\p_\nu\bn\cdot\p^\nu\bn)\p^2n_a
+(\p_\mu\p_\nu\bn\cdot\p^\nu\bn)\p^\mu n_a &\non
-(\p^2\bn\cdot\p_\mu\bn)\p^\mu n_a
-(\p_\mu\bn\cdot\p_\nu\bn)\p^\mu\p^\nu n_a
-f(n_0)f'(n_0)(\p_\mu\bn\cdot\p^\mu\bn)\delta_{a0}
+m^2\delta_{a0}&=0,
\label{eq:eom_pde}
\end{align}
where $\delta_{ab}$ is Kronecker's delta and vanishes unless $a=b$.
$n_\star$ must be strictly smaller than 1 (in the near-BPS case) and
therefore $f(n_0)$ is positive definite and greater than or equal to 1,
whereas its derivative $f'(n_0)$ is negative definite for
$n_0<n_\star$ and zero otherwise.  
Since we work with the mostly positive metric signature, the kinetic
term $\p_\mu\bn\cdot\p^\mu\bn$ reduces to $|\p_i\bn|^2$ in the static
limit, which is positive semi-definite.
Therefore, we can see that the effect of a nonconstant $f(n_0)$ is to
increase the kinetic term in the equation of motion (the first term in
the above equation) and increase the effective mass term, inside the
soliton (i.e.~where $n_0<n_\star$).

For the 1-Skyrmion, we can utilize the hedgehog Ansatz
\eqref{eq:Uhedgehog} for which the equations of motion reduce to 
\begin{align}
f^2\left(\xi_{rr} + \frac{2}{r}\xi_r - \frac{1}{r^2}\sin2\xi\right)
-ff_{n_0}\left(\sin(\xi)\xi_r^2 - \frac{2}{r^2}\sin^3\xi\right)
+\frac{2}{r^2}\sin^2(\xi)\xi_{rr} &\non
\mathop+\frac{1}{r^2}\sin(2\xi)\xi_r^2 
-\frac{1}{r^4}\sin^2(\xi)\sin(2\xi)
-m^2\sin\xi &= 0,
\label{eq:eom_ode}
\end{align}
with $f=f(n_0)=f(\cos\xi)$, $f_{n_0}=f'(n_0)=f'(\cos\xi)$ and we have
switched to PDE notation with $\xi_{r}\equiv\frac{\p\xi}{\p r}$ and so
on. 

\subsection{Inter-Skyrmion forces}\label{sec:forces}

In order to find the forces between two well separated Skyrmions in
the model at hand, we follow the analysis of Schroers
\cite{Schroers:1993yk}, adapted to include the pion mass, as in
ref.~\cite{Gudnason:2020arj}.
The first step is to linearize the ODE \eqref{eq:eom_ode} for the
1-Skyrmion:
\begin{align}
  f^2|\left(\xi_{rr} + \frac2r\xi_r - \frac{2\xi}{r^2}\right)
  -m^2\xi = 0,
  \label{eq:EOM_lin}
\end{align}
where $f^2|$ denotes the limit
\beq
f^2| := \lim_{\xi\to0} f^2(\cos\xi) =
\begin{cases}
  0, & n_\star = 1,\\
  1, & n_\star < 1.
\end{cases}
\eeq
Since we have excluded the BPS limit, we have $n_\star<1$ and thus
$f^2|=1$.
The solution to the linearized EOM for the 1-Skyrmion
\eqref{eq:EOM_lin} is thus given by 
\beq
\xi = \frac{q m}{4\pi} k_1(mr),
\eeq
where $q$ is a positive constant, $4\pi/m$ is a convenient
normalization and $k_n$ is the spherical modified Bessel function of
the second kind, which has the asymptotic behavior 
\beq
\xi = -q\frac{\d}{\d r}\left(\frac{e^{-m r}}{4\pi m r}\right),
\eeq
where we have used a recursion relation between the Bessel functions
and that $k_0(x)=e^{-x}/x$.
It is now elementary to show that the linearized Skyrme field at
asymptotic distances is given by three orthogonal dipoles
\beq
U = \mathbf{1}_2 + \i\bpi\cdot\bsigma, \qquad
\bpi = (\pi_1,\pi_2,\pi_3), 
\eeq
with
\beq
\bpi = -q\frac{\p}{\p\bx}\left(\frac{e^{-m r}}{4\pi m r}\right),
\label{eq:3dipoles}
\eeq
and the corresponding quadratic static Lagrangian density is given by 
\beq
\Lag_{\rm quad} = -\p_i\bpi\cdot\p_i\bpi - m^2\bpi\cdot\bpi.
\label{eq:Lquad}
\eeq
The solution \eqref{eq:3dipoles} is, to leading order in $1/r$, the
solution to the equation of motion for the quadratic Lagrangian
\eqref{eq:Lquad} with a point source of charge $q$
\beq
(\p_i^2 - m^2) \bpi = -q\delta^{(3)}(\bx),
\eeq
whose exact solution is
\beq
\bpi = -\frac{q}{4\pi m r}\frac{\p}{\p\bx}e^{-m r}.
\eeq
Placing two Skyrmions at $\bX^{(1),(2)}$ with orientation
$\mathscr{R}_{ab}^{(1),(2)}$, where the latter are SO(3) rotation
matrices, we have
\begin{equation}
\pi_a^{(1)} = -q\mathscr{R}_{ab}^{(1)}\frac{\p}{\p x^b}
\left(\frac{e^{-m|\bx-\bX^{(1)}|}}{4\pi m |\bx-\bX^{(1)}|}\right),\qquad
\pi_a^{(2)} = -q\mathscr{R}_{ab}^{(2)}\frac{\p}{\p x^b}
\left(\frac{e^{-m|\bx-\bX^{(2)}|}}{4\pi m |\bx-\bX^{(2)}|}\right).
\end{equation}
The interaction potential can straightforwardly be calculated
following refs.~\cite{Schroers:1993yk,Gudnason:2020arj}, yielding 
\beq
V_{\rm int} \simeq \frac{q^2e^{-mR}}{16\pi R} \widehat{\bR}\cdot\mathscr{O}\widehat{\bR},
\label{eq:Vint}
\eeq
where we have defined $\bR\equiv|\bX^{(1)}-\bX^{(2)}|$,
$R\equiv|\bR|$,
$\mathscr{O}\equiv(\mathscr{R}^{(1)})^{\rm T}\mathscr{R}^{(2)}$,
and $\widehat{\bR}\equiv\bR/R$.

The interaction potential \eqref{eq:Vint}, which is valid for
$n_\star<1$, shows that two Skyrmions at a separation distance $R$
in the \emph{same} orientation, are mutually repulsive, whereas two
Skyrmions in the \emph{opposite} orientation, are mutually
attractive.
By opposite orientation, we mean that
$\widehat{\bR}\cdot\mathscr{O}\widehat{\bR}<0$, which can be obtained
by rotating the second Skyrmion by 180 degrees about an axis
perpendicular to the axis joining their centers.

To summarize, we have thus shown that for $n_\star=1$, the kinetic
term is turned off at asymptotic distances, and hence only the Skyrme
term and the mass term remain.
Since the kinetic term is the origin of the fact that the Skyrmion can
be viewed from afar as three orthogonal dipoles, they are the reason for
the attraction between the Skyrmions at asymptotic distances (if they
are rotated into the attractive channel, see the previous paragraph).
To linear order, the Skyrme term does not play a role and hence we
have not explicitly shown that the Skyrmions become repulsive in the
BPS limit (i.e.~$n_\star=1$).
But since that limit is unphysical, we will not consider such
investigation here.

\section{Numerical results}\label{sec:num}

In this section we will solve the equations of motion
\eqref{eq:eom_pde} numerically. The method we will use is based on a 
fourth-order finite-difference scheme with a 5-point stencil for the
field derivatives and as the algorithm to find the minimizer of the energy
functional from given initial data, we will use the arrested Newton
flow described in ref.~\cite{Gudnason:2020arj}.
The lattices used in this paper are cubic lattices with $120^3$--$240^3$
lattice points and lattice spacing $h_x\leq 0.15$.
The estimated numerical error is less than $10^{-6}$. 

As initial data, we will use the rational map approximations (RMA)
of ref.~\cite{Houghton:1997kg} for the Skyrmions with topological
degrees 1 through 8.
The $B=8$ solution obtained from the RMA has dihedral symmetry and we
shall denote it as the $B=8_h$ solution.
We will use two further initial conditions for the $B=8$ sector, which
are composed by two $B=4$ cubes placed in the vicinity of one another:
one that is a translated copy of the first, which we shall call the
$B=8_u$ for the untwisted chain and the other is the twisted
chain, $B=8_t$, obtained by rotating one of the two cubes by 90
degrees about the axis joining their centers. 
The initial condition for the two cubes is prepared by means of the
asymmetric product Ansatz
\beq
U_8 = U_4 U_4',
\eeq
with $U_4$ one cube made by the RMA for the $B=4$ Skyrmion, translated
by a bit more than half the cube's size in the $-x$ direction and the
other cube, $U_4'$, is translated by the same distance in the $+x$
direction.  

A comment in store is about the cusp, present in the function $f$ of
eq.~\eqref{eq:fe_funs}.
That is, the function $f$ is not smooth (as its derivative jumps from
a negative value to zero at $n_0=n_\star$) and a smooth interpolation
should be invented.
Then the lattice spacing of the numerical lattice should be small
enough to resolve the invented smooth interpolation.
However, since this is an arbitrariness in the near-BPS version of the
model, we chose to deal with this issue very pragmatically, namely to
let the discrete derivatives smooth out the cusp.
To make sure that it is done consistently in all solutions, we have
used the same lattice spacing for all solutions of a given $n_\star$.

\begin{figure}[!htp]
  \begin{center}
    \mbox{\subfloat[$B=1$]{\includegraphics[width=\linewidth]{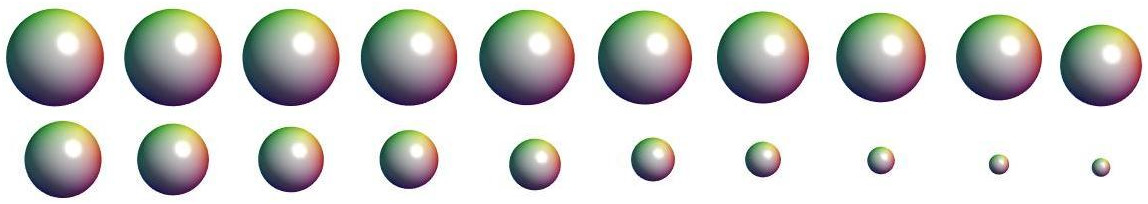}}}
    \mbox{\subfloat[$B=2$]{\includegraphics[width=\linewidth]{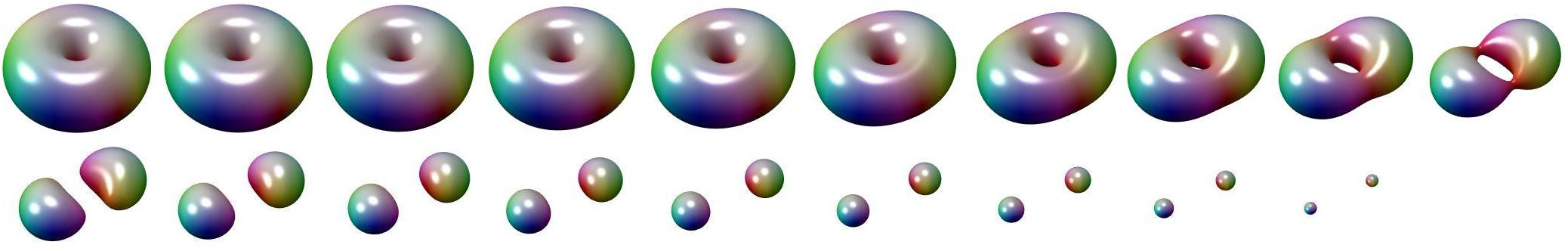}}}
    \mbox{\subfloat[$B=3$]{\includegraphics[width=\linewidth]{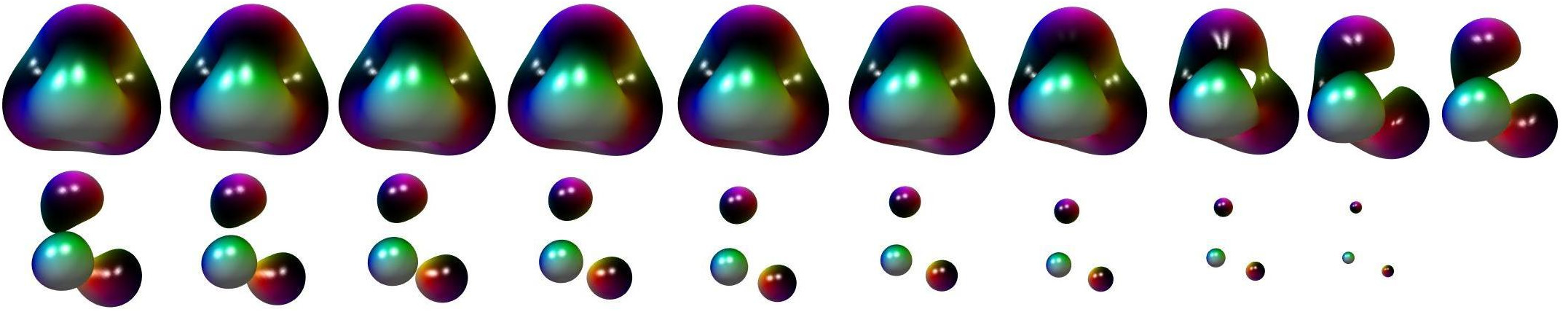}}}
    \mbox{\subfloat[$B=4$]{\includegraphics[width=\linewidth]{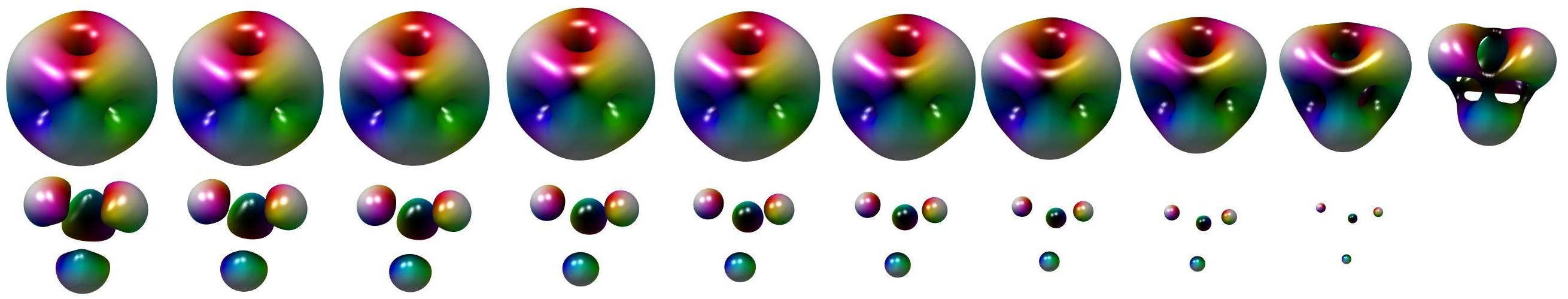}}}
    \mbox{\subfloat[$B=5$]{\includegraphics[width=\linewidth]{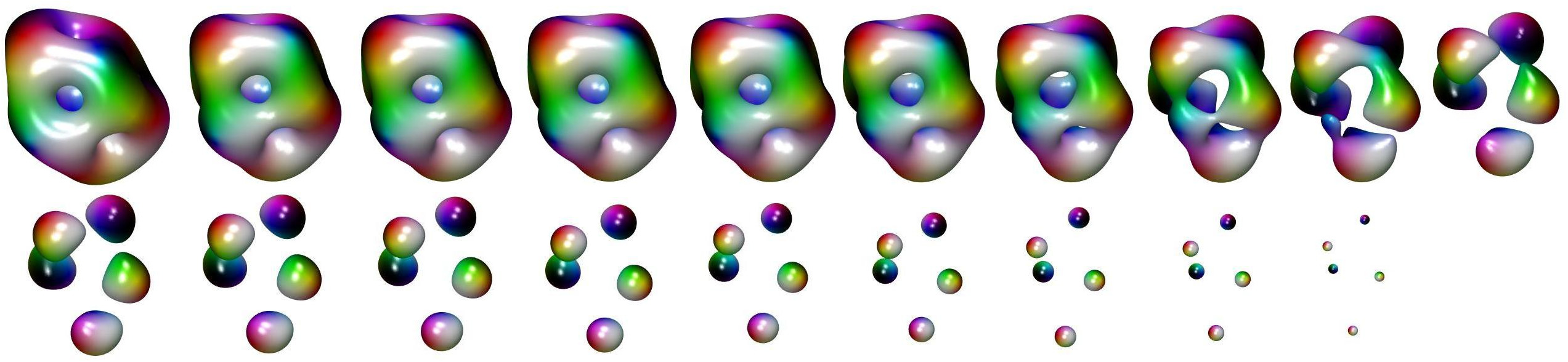}}}
    \caption{Skyrmion solutions for (a) $B=1$ to (e) $B=5$. Each
      panel shows 19 figures, except for (a) which shows 20 figures,
      which are isosurfaces of the topological charge density (TCD) at
      a quarter of its maximum 
      value, and the first row (from left to right) represents the
      solution with $n_\star$ increasing from $-1$ to $-0.1$ in steps
      of $0.1$ and in the second row $n_\star$ increases from $0$ to
      $0.8$ (and $0.9$ in panel (a)), again in steps of $0.1$.
      The color scheme is described in the text. 
    }
    \label{fig:mon1-5}
  \end{center}
\end{figure}

\begin{figure}[!htp]
  \begin{center}
    \mbox{\subfloat[$B=6$]{\includegraphics[width=\linewidth]{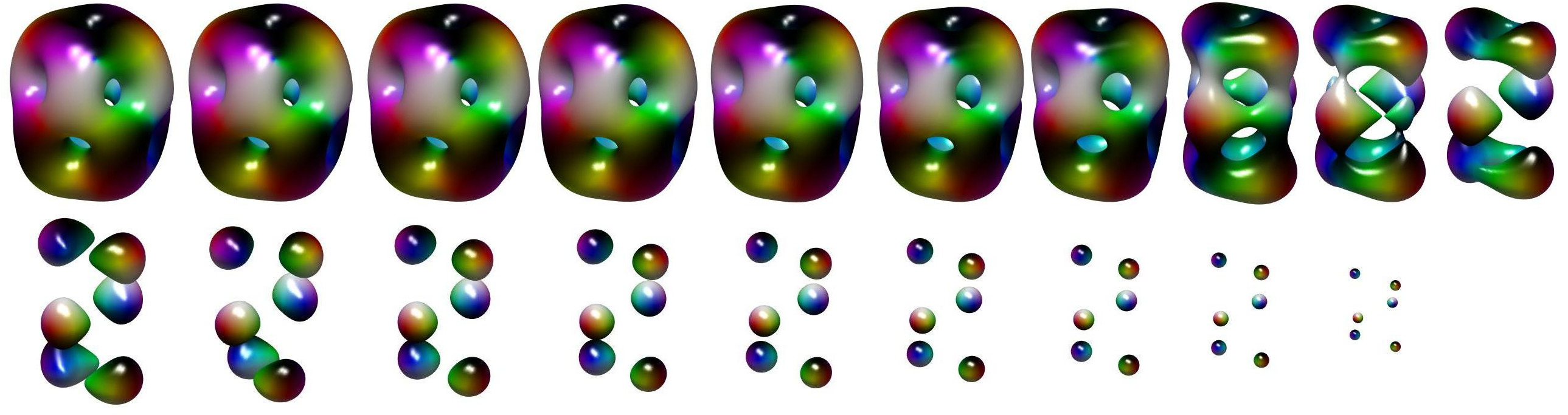}}}
    \mbox{\subfloat[$B=7$]{\includegraphics[width=\linewidth]{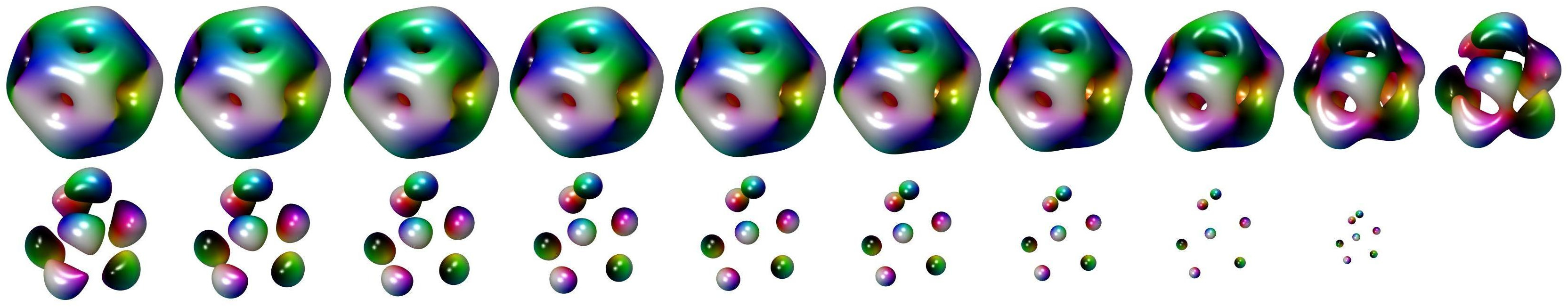}}}
    \mbox{\subfloat[$B=8_h$]{\includegraphics[width=\linewidth]{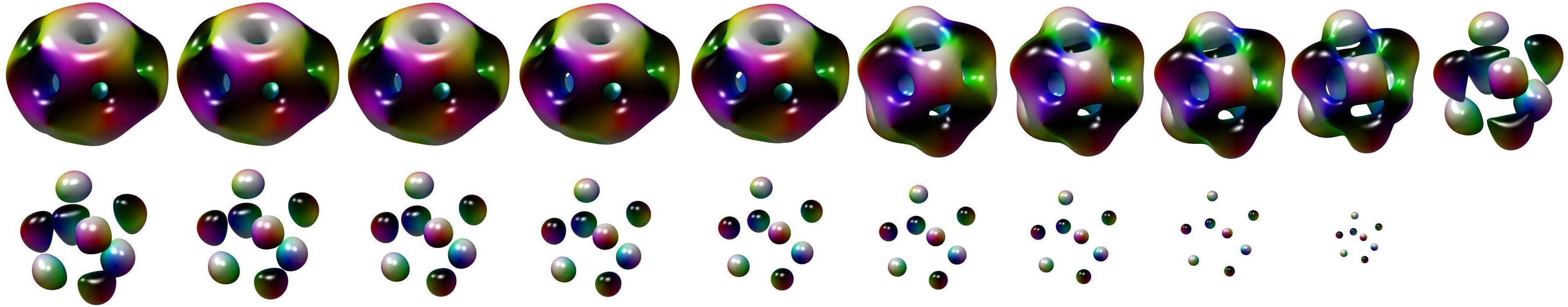}}}
    \mbox{\subfloat[$B=8_u$]{\includegraphics[width=\linewidth]{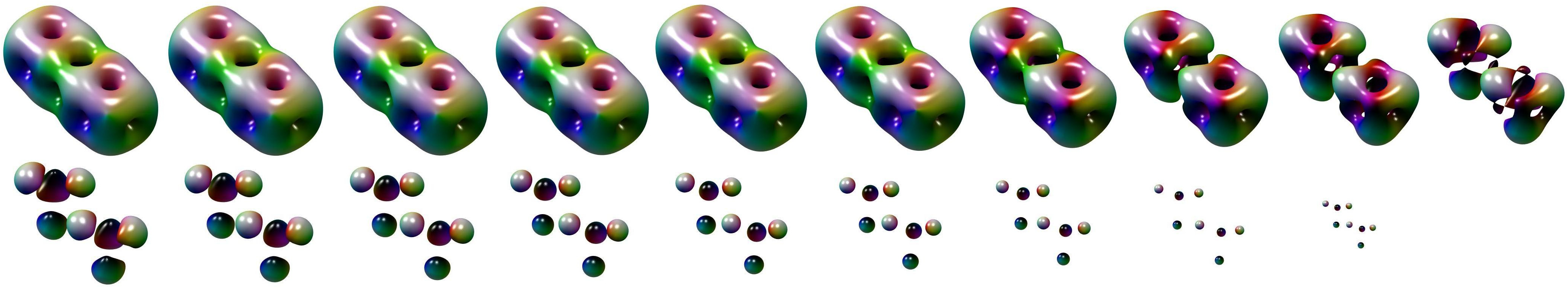}}}
    \caption{Skyrmion solutions for (a) $B=1$ to (e) $B=5$. Each
      panel shows 19 figures, which are isosurfaces of the TCD at a
      quarter of its maximum value, and the first row (from left to
      right) represents the solution with $n_\star$ increasing from
      $-1$ to $-0.1$ in steps of $0.1$ and in the second row $n_\star$
      increases from $0$ to $0.8$, again in steps of $0.1$.
      The color scheme is described in the text.
    }
    \label{fig:mon6-8u}
  \end{center}
\end{figure}

We are now ready to present the numerical solutions for the near-BPS
dielectric Skyrme model for various values of $n_\star$, starting
with the potential turned off ($m=0$).
In this massless version of the model, it is clear that $n_\star$
interpolates between the standard massless Skyrme model and a BPS
theory.
Furthermore, as mentioned above, the BPS theory does not possess
multi-Skyrmions -- they are unbound.
This means that, even though we exclude the point $n_\star=1$, we
should be able to reach arbitrarily small binding energies by cranking
up $n_\star$.
The question is what happens to the multi-Skyrmions, which we shall
now answer with figs.~\ref{fig:mon1-5} and \ref{fig:mon6-8u}.
The figures show arrays of solutions in the form of isosurfaces of the
topological charge density (TCD) at a fixed level set, which is taken to be
a quarter of the maximum TCD of the given solution.
We furthermore color in the isosurface using a standard coloring
scheme based on the normalized pion 3-vector
\beq
\hat\bpi = \frac{(n_1,n_2,n_3)}{\sqrt{n_1^2+n_2^2+n_3^2}},
\eeq
such that $\hat\pi_3=1$ is white, $\hat\pi_3=-1$ is black and
$\hat\pi_3=0$ is a color determined by
$\hat\pi_1+\i\hat\pi_2=e^{\i H}$, where $H$ is the hue of the color.
In particular $H=0$ is red, $H=2\pi/3$ is green and $H=4\pi/3$ is
blue.
Each panel in the figures presenting Skyrmion solutions by isosurfaces
of their TCD are made of 19 (20) figures, with the first row showing
$n_\star=-1,-0.9,\ldots,-0.1$ and the second row showing
$n_\star=0,0.1,\ldots,0.8$ for baryon numbers $B\geq 2$
($n_\star=0,0.1,\ldots,0.9$ for $B=1$). 

\begin{figure}[!htp]
  \begin{center}
    \mbox{\subfloat[$m=0$]{\includegraphics[width=0.49\linewidth]{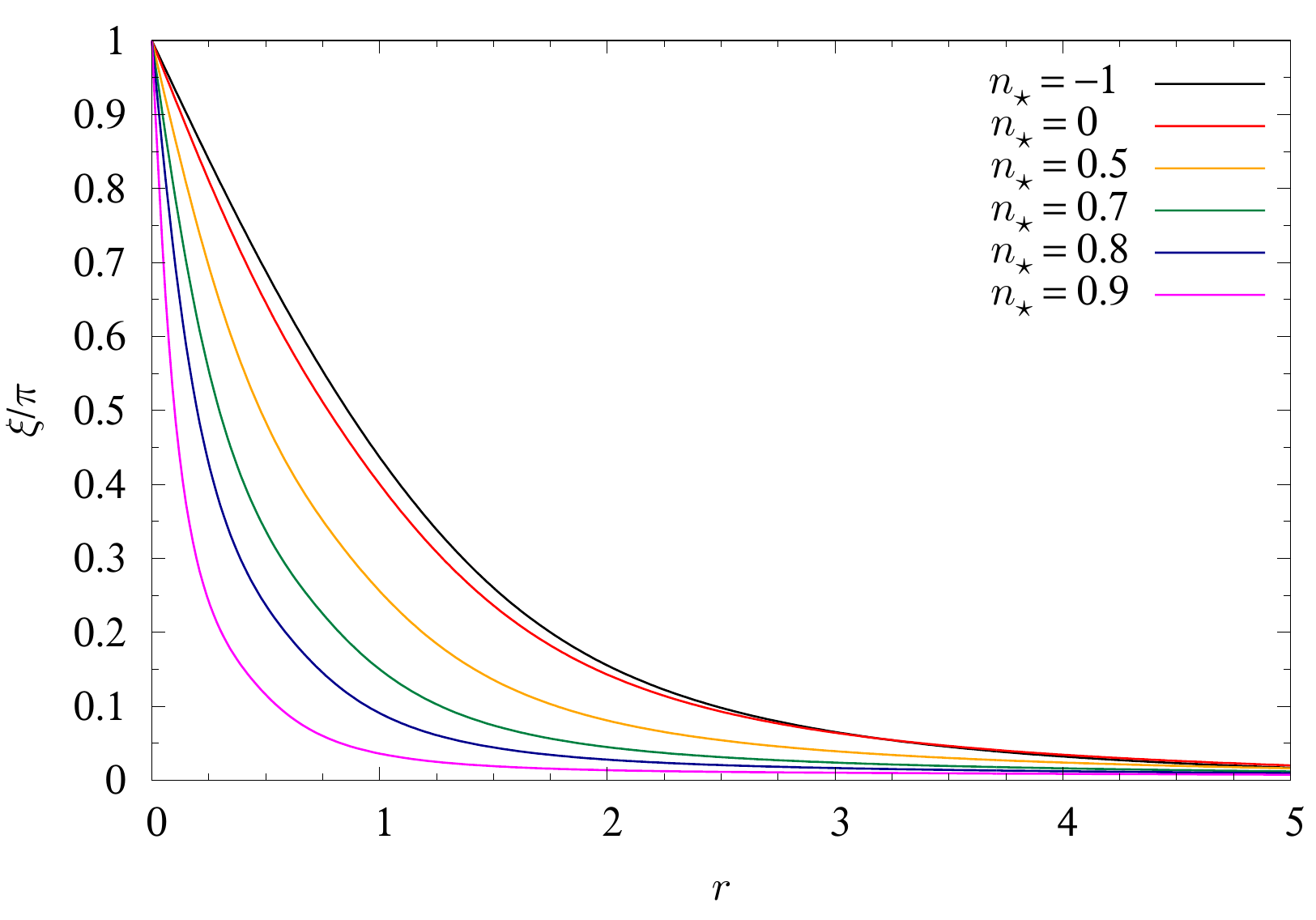}}
      \subfloat[$m=1$]{\includegraphics[width=0.49\linewidth]{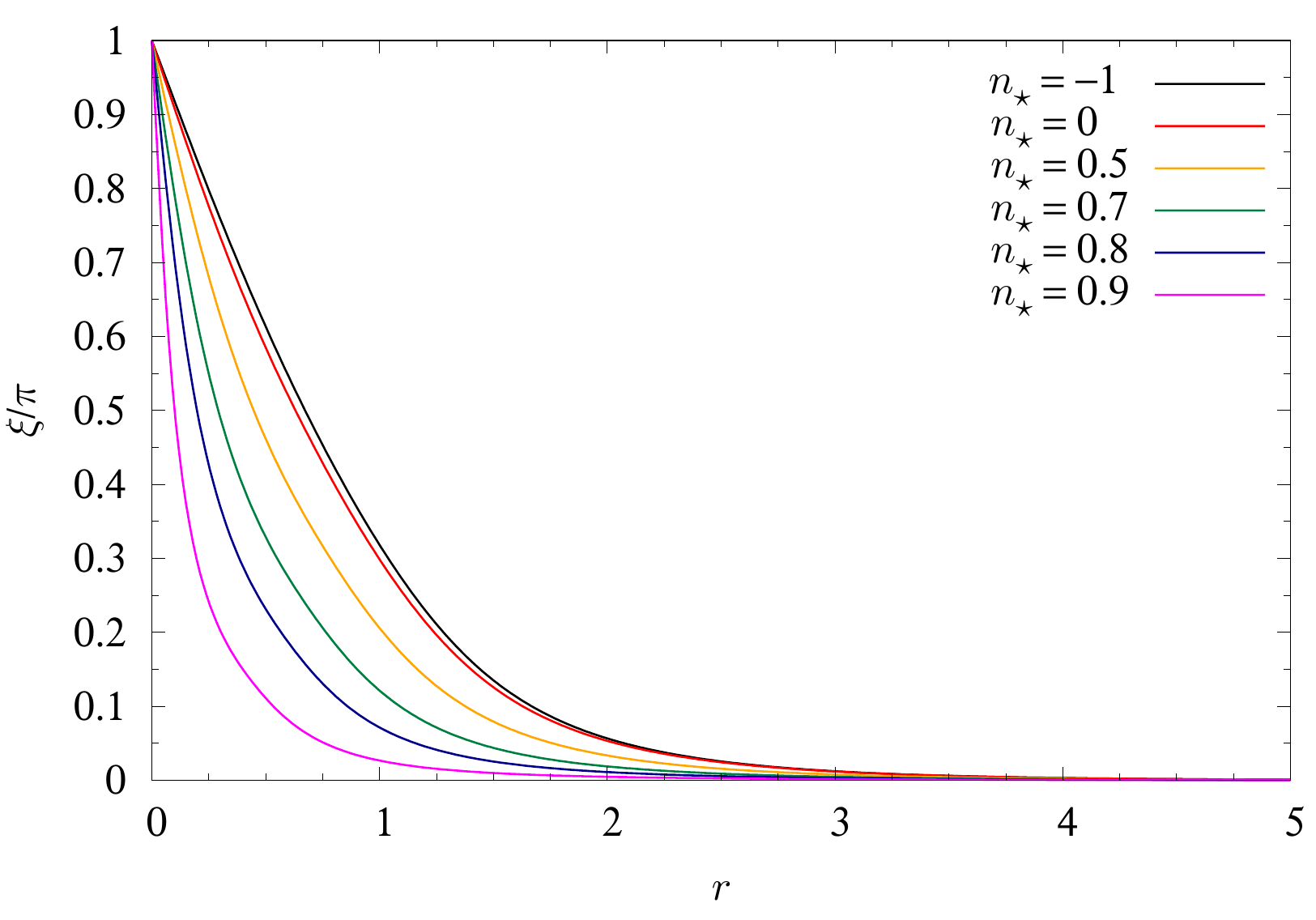}}}
    \caption{$B=1$ Skyrmion solutions to the ODE \eqref{eq:eom_ode},
      for various values of $n_\star=-1,0,0.5,0.7,0.8,0.9$.
    } 
    \label{fig:B1}
  \end{center}
\end{figure}

The 1-Skyrmion retains its spherical symmetry for all values of
$n_\star$, see fig.~\ref{fig:mon1-5}(a).
Although we have divided $f$ by $1-n_\star$, the Skyrmions still
shrink for $n_\star\geq 0$ (if we had not divided by $1-n_\star$, $f$
would become very large for $n_\star$ approaching 1).
Instead now, the tail of the Skyrmion can be captured with the same
size lattices, but the BPS nature of the solutions, make them quite
peaked at the origin, see fig.~\ref{fig:B1}. 
For that reason, we have gradually shrunk the lattice spacing for
large $n_\star$, but not so much that the tails of the solutions would
not be contained.
In all solutions, the topological charge calculated from the TCD is
accurate to the $10^{-4}$ level or better.

As already mentioned, the massless near-BPS dielectric Skyrme model is
exactly the standard massless Skyrme model for $n_\star=-1$ and thus
the solutions coincide in that limit.
The 2-Skyrmion is shown in fig.~\ref{fig:mon1-5}(b) and starts out 
with a torus shape (top left) and is gradually transformed into two
separate 1-Skyrmions as $n_\star$ tends to 0 (top row). $n_\star=0$ is
the Skyrmion shown in the left bottom of panel (b).
As $n_\star$ is further increased to $0.8$ the two 1-Skyrmions become
smaller and mutually less interacting.
The weak interaction between the two 1-Skyrmions by means of just a
mere overlap of their respective soliton tails is the reason for the
small binding energy.

Let us summarize the solutions of the remaining three topological
sectors in fig.~\ref{fig:mon1-5}.
The 3-Skyrmion starts with tetrahedral symmetry for $n_\star=-1$ and
deforms into three 1-Skyrmions placed at the vertices of a triangle.
The 4-Skyrmion starts off with octahedral (cubic) symmetry as a
platonic solid for $n_\star=-1$ and deforms into a tetrahedrally
symmetric soliton for $n_\star\simeq-0.2$ and then gradually dissolves
into four 1-Skyrmions placed at the vertices of a tetrahedron.
The 5-Skyrmion begins with dihedral symmetry and gradually deforms
into five 1-Skyrmions placed approximately on the vertices of a
face-centered cubic (FCC) lattice with four of them at the vertices of
a tetrahedron and the last one a satellite. 

Before discussing the emerged pattern, let us summarize also the
Skyrmion solutions in the topological charge sectors $B=6$ through
$B=8$, shown in fig.~\ref{fig:mon6-8u}.
For $n_\star=-1$ the 6-Skyrmion has dihedral symmetry
(fig.~\ref{fig:mon6-8u}(a)), the 7-Skyrmion has icosahedral symmetry
(fig.~\ref{fig:mon6-8u}(b)), the $B=8_h$ has dihedral symmetry
(fig.~\ref{fig:mon6-8u}(c)) and the $B=8_u$ has cubic symmetry
(fig.~\ref{fig:mon6-8u}(d)) \cite{Battye:2006na,Gudnason:2018ysx}.
Of the two $B=8$ Skyrmion solutions, the $B=8_h$ solution with
dihedral symmetry has the lowest energy in the standard massless
Skyrme model and this is indeed the prediction of the RMA
\cite{Battye:1997qq}. 
Upon increasing $n_\star$ they all deform into $B$ 1-Skyrmions placed
at vertices of an FCC-lattice.

A clear pattern has thus emerged and in the near-BPS limit
(i.e.~$n_\star$ tending towards 1), the Skyrmions become point-particle
Skyrmions placed at the vertices of an FCC-lattice.
This is exactly the kind of solution also possessed by the lightly
bound Skyrme model \cite{Gillard:2015eia,Gillard:2016esy}, which is
simply the standard (massive) Skyrme model with the addition of the
potential
\beq
m_4^2(1 - n_0)^4.
\eeq
The same type of solution also exists for the loosely bound Skyrme
model, with the above potential replaced by
\beq
m_2^2(1 - n_0)^2,
\eeq
see refs.~\cite{Gudnason:2016mms,Gudnason:2016cdo,Gudnason:2018jia}.
Furthermore, in the context of holographic QCD or specifically the
Sakai-Sugimoto model, the point-particle Skyrmion solutions appear at
low energies as a dimensional reduction of point-particle instantons
(i.e.~instantons with small size moduli) in the limit of large 't
Hooft coupling \cite{Baldino:2017mqq}.

Before considering the energies of the Skyrmion solutions, we will
include the case of the \emph{massive} near-BPS dielectric Skyrme
model, viz.~turning on $m=1$ and recalculate the solutions.
Because the solutions are quite similar and the deformation to
point-particle solutions as 1-Skyrmions placed at the vertices of an
FCC-lattice is qualitatively the same, we will not show the analogues
of figs.~\ref{fig:mon1-5} and \ref{fig:mon6-8u} for $m=1$.

\begin{figure}[!htp]
  \begin{center}
    \mbox{\subfloat[$B=8_t$]{\includegraphics[width=\linewidth]{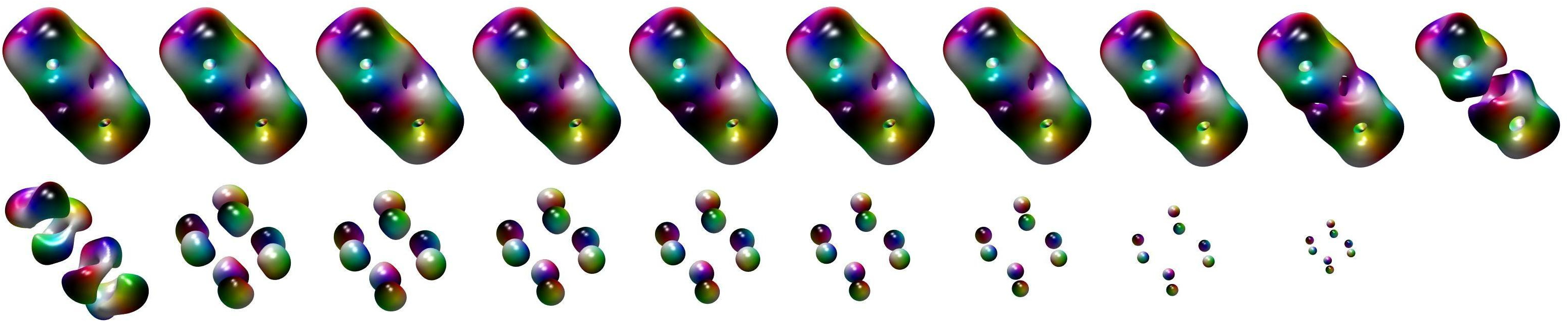}}}
    \caption{Skyrmion solutions for the $B=8_t$ twisted chain
      Skyrmion. Each panel shows 19 figures, which are isosurfaces of
      the TCD at a quarter of its maximum value, and the first row
      (from left to right) represents the solution with $n_\star$
      increasing from $-1$ to $-0.1$ in steps of $0.1$ and in the second
      row $n_\star$ increases from $0$ to $0.8$, again in steps of
      $0.1$.
      The color scheme is described in the text. }
    \label{fig:mon8tm1}
  \end{center}
\end{figure}

One qualitative difference does manifest itself already for the
standard Skyrme model (i.e.~at $n_\star=-1$).
That is, the dihedrally symmetric $B=8_h$ Skyrmion is slightly lifted
in energy and in particular a twisted chain, $B=8_t$, made of two
$B=4$ cubes appears as a solution, see fig.~\ref{fig:mon8tm1}.
The twisted chain, $B=8_t$, is obtained from the \emph{untwisted}
chain, $B=8_u$ by rotating one of the cubes by 90 degrees about the
axis joining their centers.
The twisted chain, $B=8_t$, does not exist in the massless case
($m=0$), whereas it does exist in the massive case ($m=1$)
\cite{Battye:2006na}. 
Attempting to find the twisted chain ($B=8_t$) in the massless case,
results in a relatively fast decay into the dihedrally symmetric
solution ($B=8_h$), which can be pictured by inflating the middle of 
the chained solution.

\begin{figure}[!htp]
  \begin{center}
    \mbox{\subfloat[$m=0$]{\includegraphics[width=0.49\linewidth]{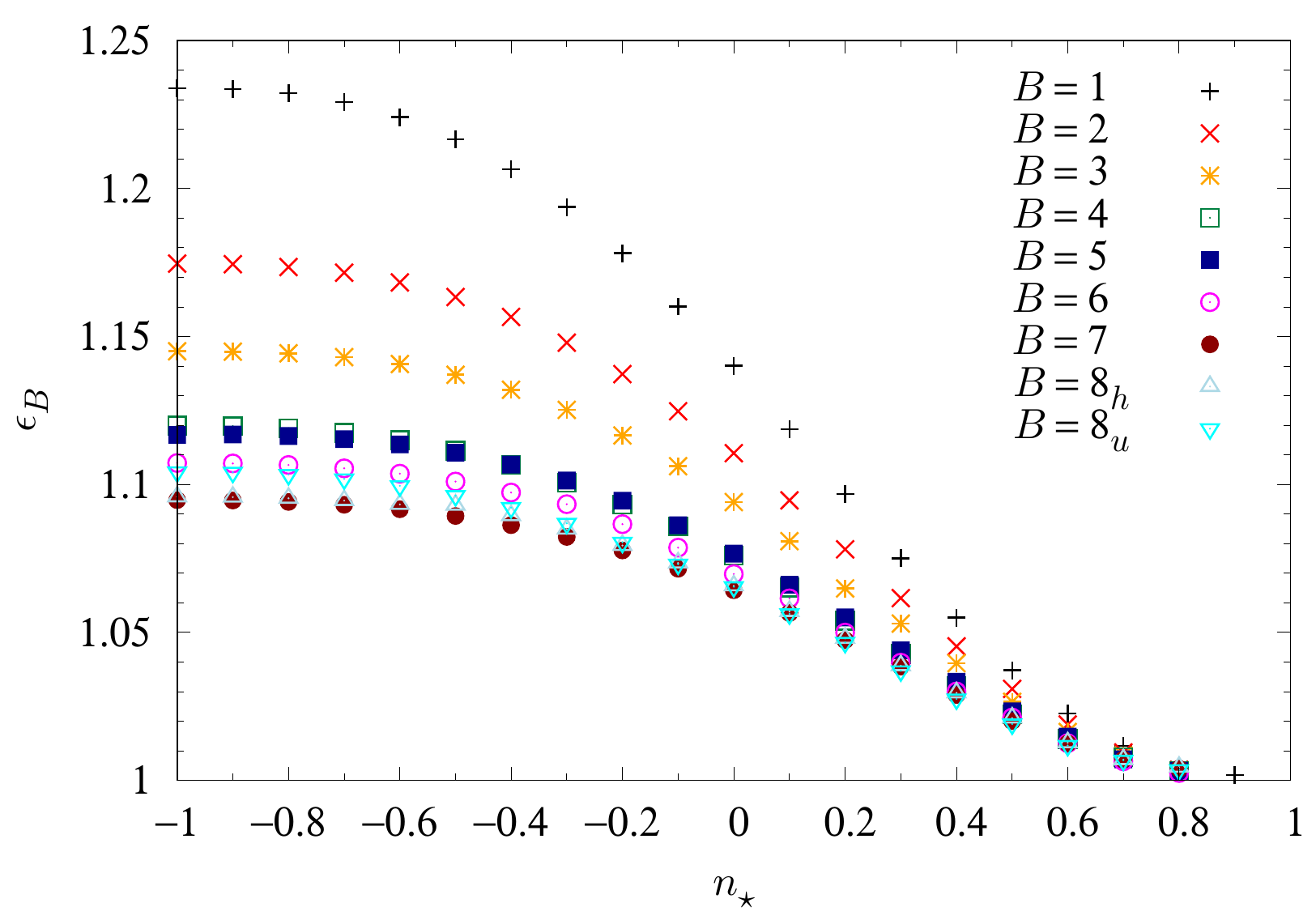}}
      \subfloat[$m=1$]{\includegraphics[width=0.49\linewidth]{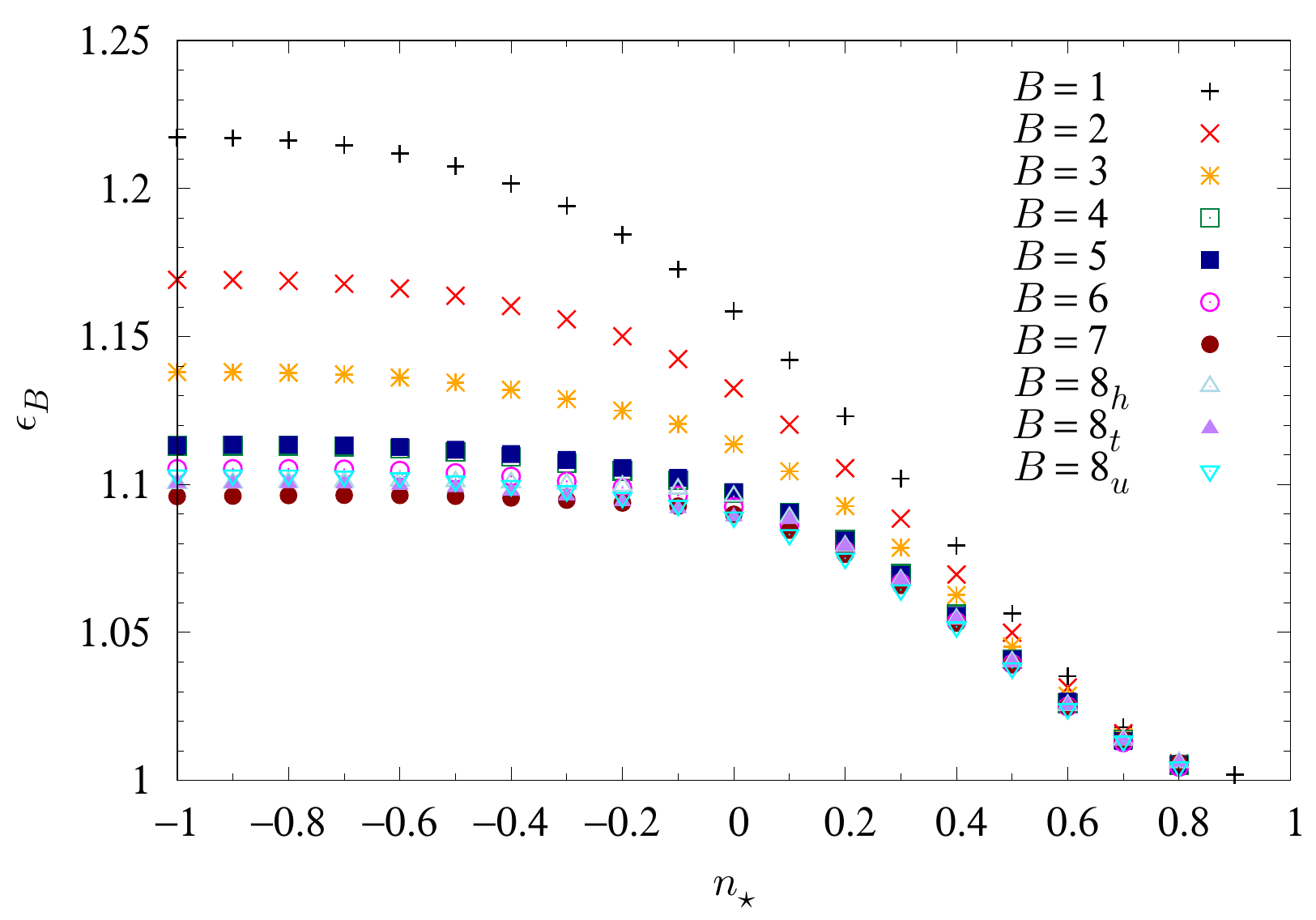}}}
    \caption{Energy of all \emph{stable} Skyrmion solutions normalized
      by their respective energy bounds, $\epsilon_B$: (a)
      eq.~\eqref{eq:epsilon_B}: without the pion mass term, (b)
      eq.~\eqref{eq:epsilon_B_m1}: with the pion mass term and $m=1$.
    }
    \label{fig:en}
  \end{center}
\end{figure}

In fig.~\ref{fig:en} we show the energies normalized by their
energy bound, $\epsilon_B$, of all stable
Skyrmion solutions for $B=1$ through $B=8$, with and without the mass
term turned on. 
For the massless case ($m=0$), the energy normalized by its
Bogomol'nyi bound is given by eq.~\eqref{eq:epsilon_B}, whereas in the
massive case ($m=1$), it is given by eq.~\eqref{eq:epsilon_B_m1}.
First of all, we can see that all energies lie above their respective
energy bound, i.e.~$\epsilon_B>1$ for all calculated solutions.
Since the energies come really close to the bound for $n_\star=0.8$,
this is a signal of our accuracy being quite good.
We do not trust solutions for $n_\star>0.8$, because the 1-Skyrmions
become too peaked -- but at the same time still possessing quite long
tails -- to be captured reasonably on the lattices, with the lattice
spacings used in this paper.
Second of all, we can also see by inspection of the figure, that all
solutions have positive binding energies.

We should explain what we mean by \emph{stable} Skyrmions solutions.
For $n_\star=-1$, the Skyrmions are quite symmetric, possessing a
discrete symmetry.
It turns out that by increasing $n_\star$, such symmetry is broken and
the solution gradually transforms into a less symmetric state.
Nevertheless, as we shall see shortly, the highly symmetric solution
with the symmetries of the $n_\star=-1$ solution, continues to exist
as a metastable (or perhaps unstable) solution for higher $n_\star$.
In that sense, only the stable solutions are shown in
fig.~\ref{fig:en}, with the exception of the $B=8$ sector, which is
more complicated, see below.

\begin{figure}[!htp]
  \begin{center}
    \mbox{\subfloat[]{\includegraphics[width=0.49\linewidth]{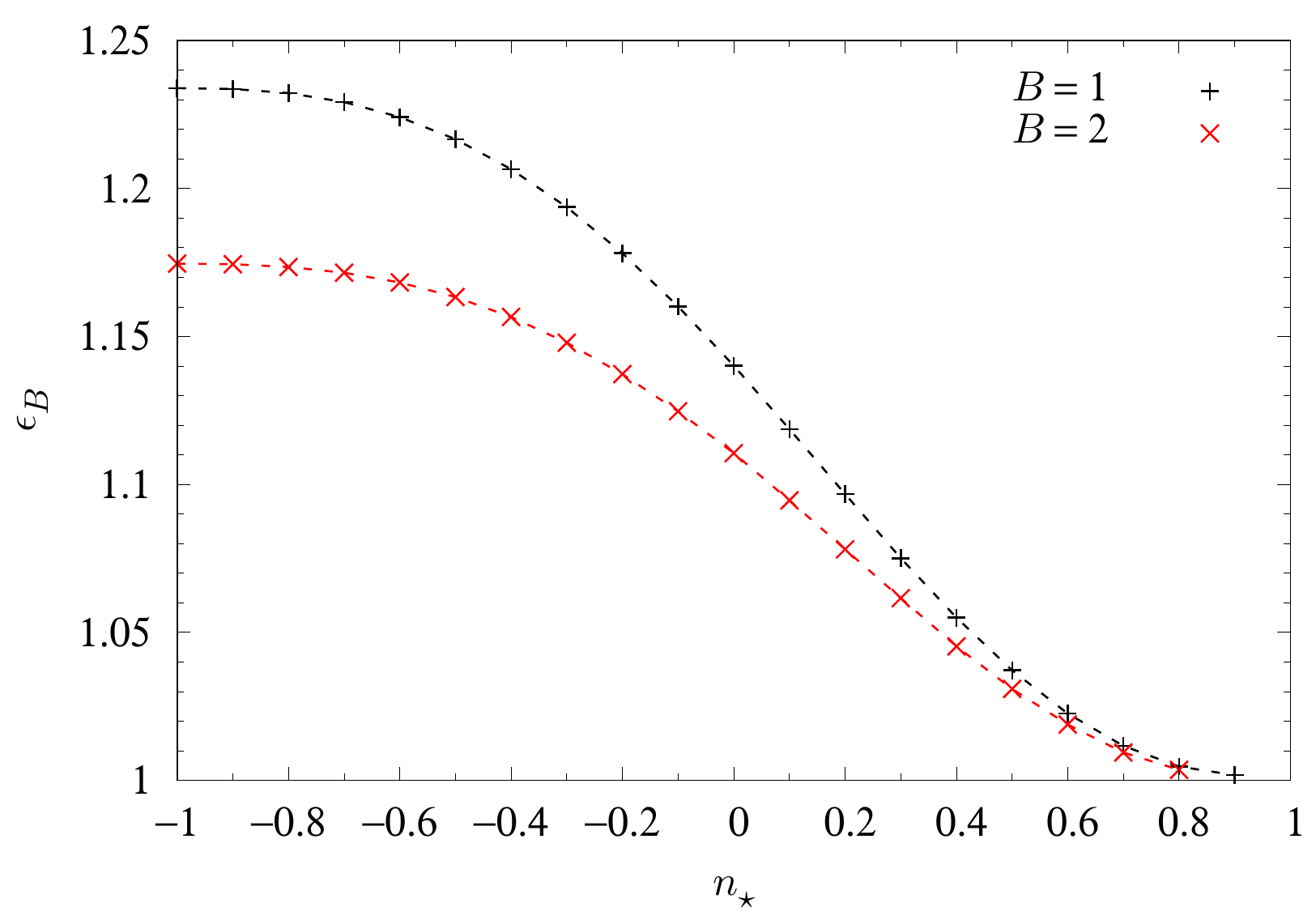}}
      \subfloat[]{\includegraphics[width=0.49\linewidth]{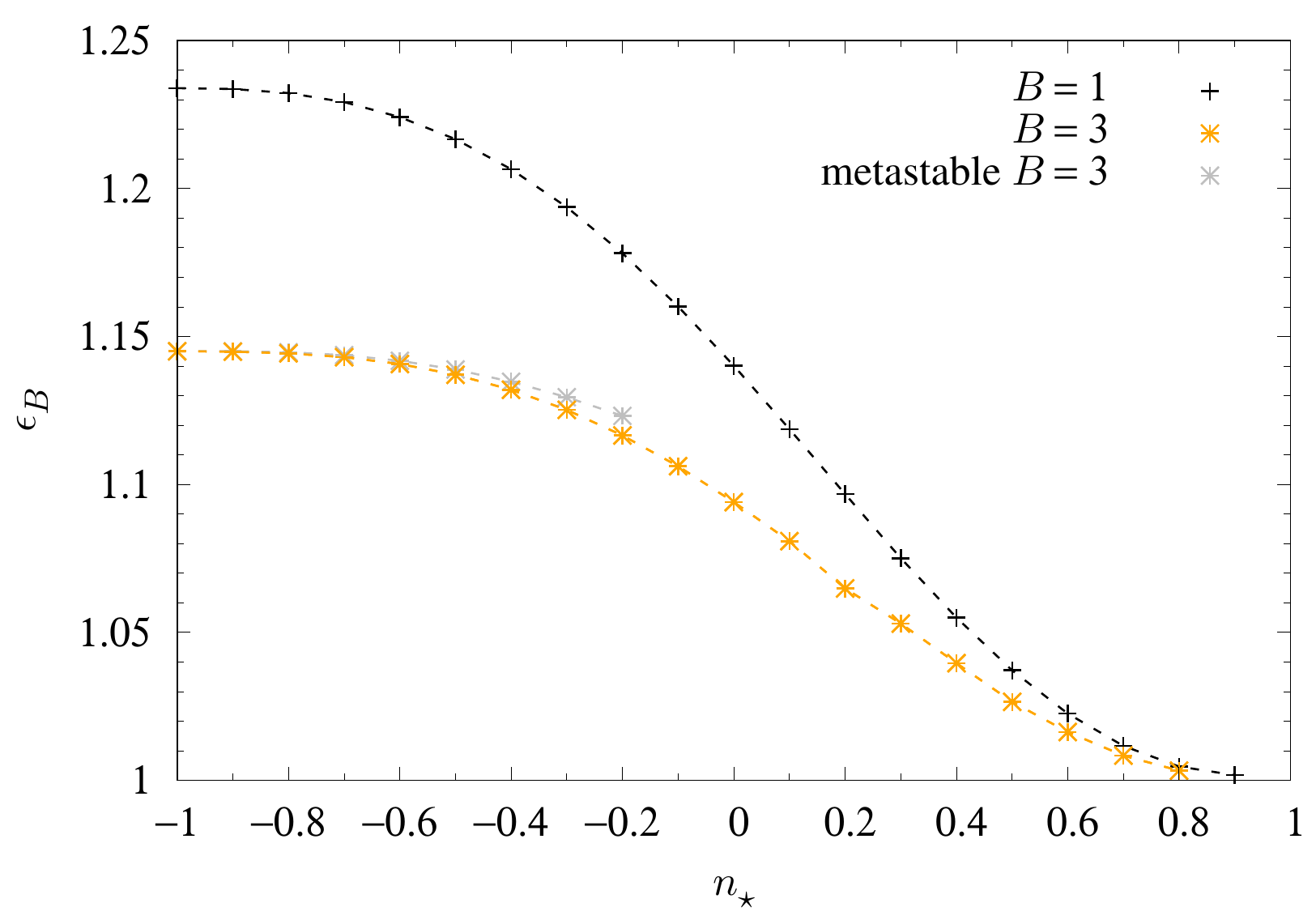}}}
    \mbox{\subfloat[]{\includegraphics[width=0.49\linewidth]{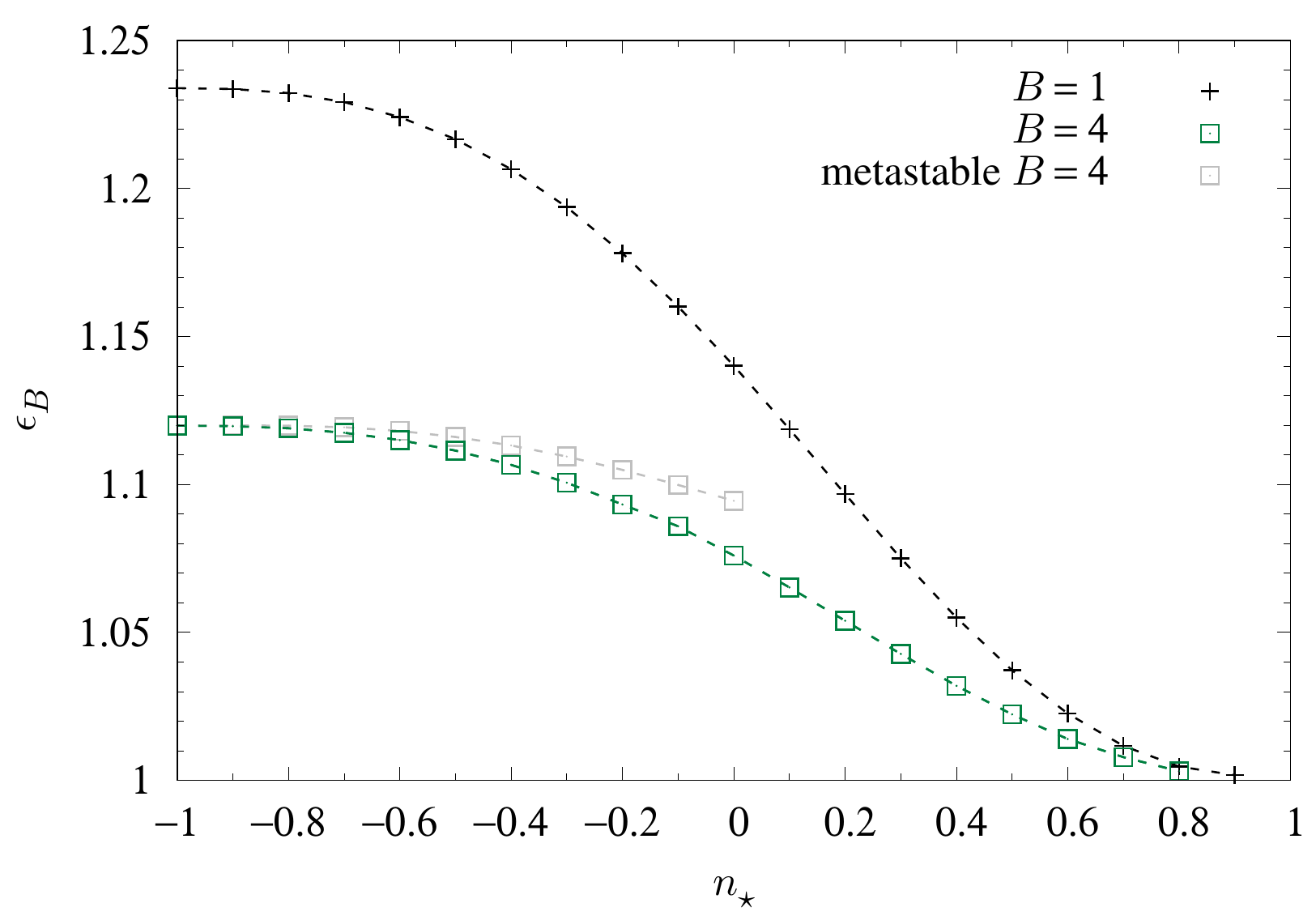}}
      \subfloat[]{\includegraphics[width=0.49\linewidth]{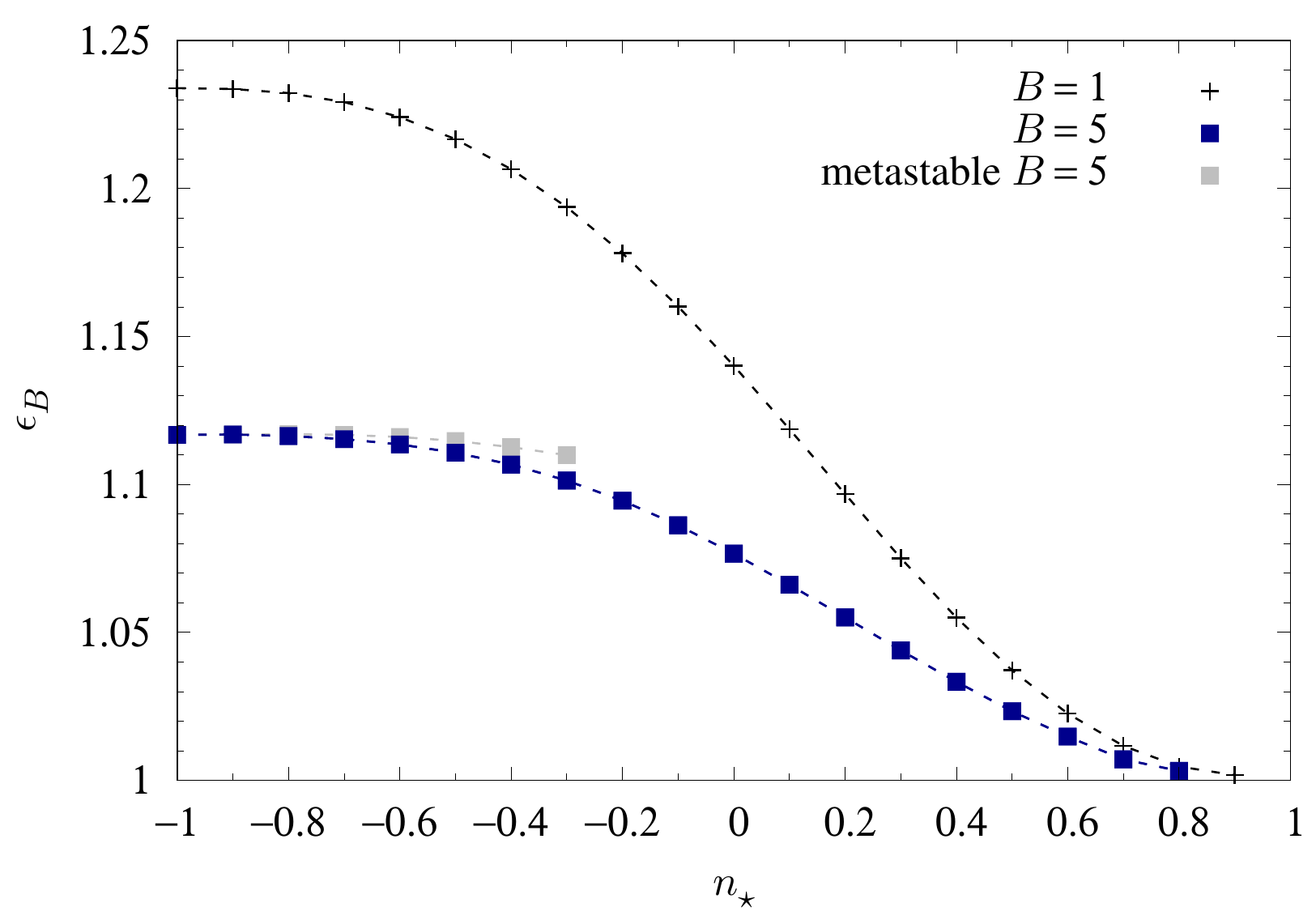}}}
    \mbox{\subfloat[]{\includegraphics[width=0.49\linewidth]{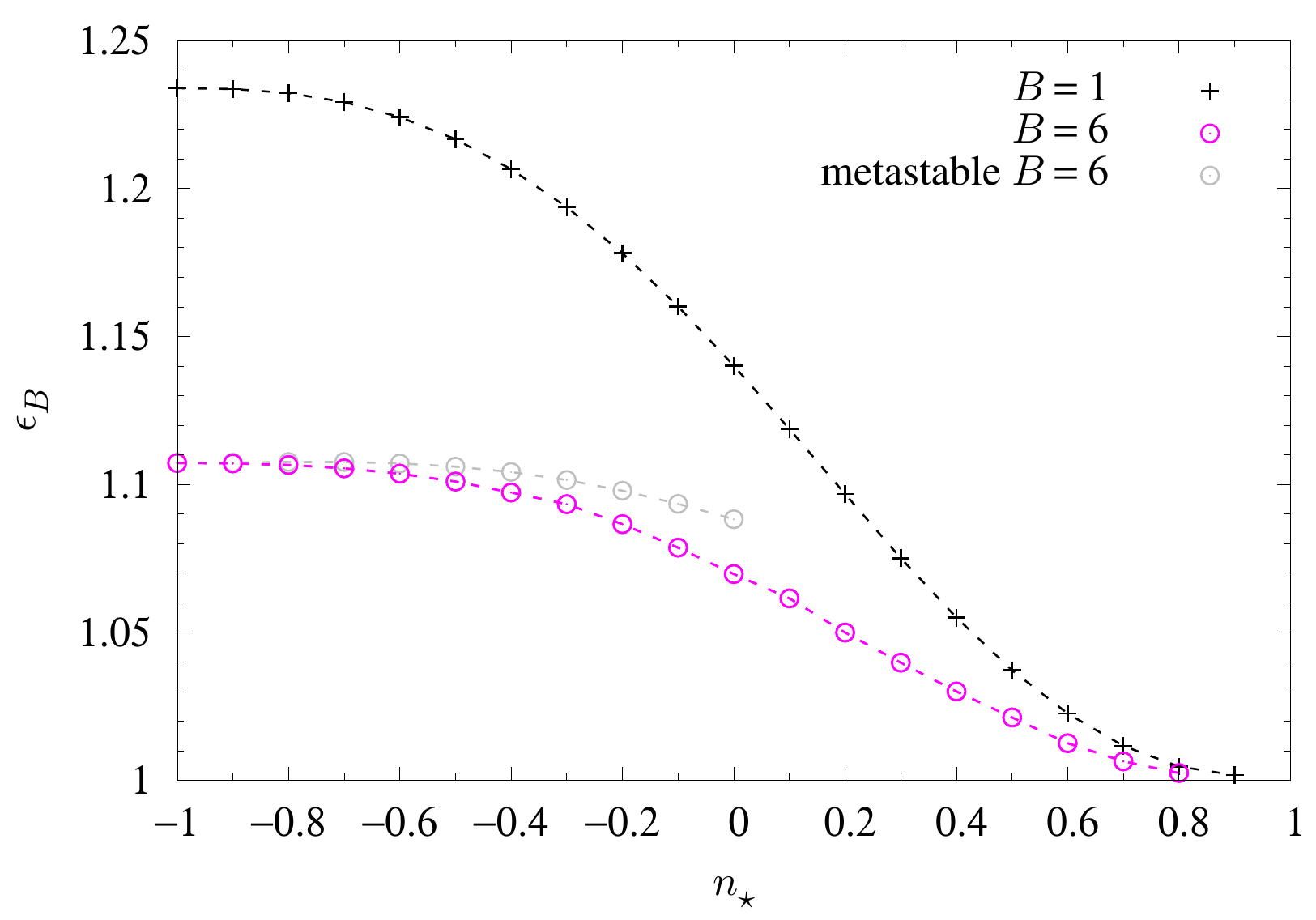}}
      \subfloat[]{\includegraphics[width=0.49\linewidth]{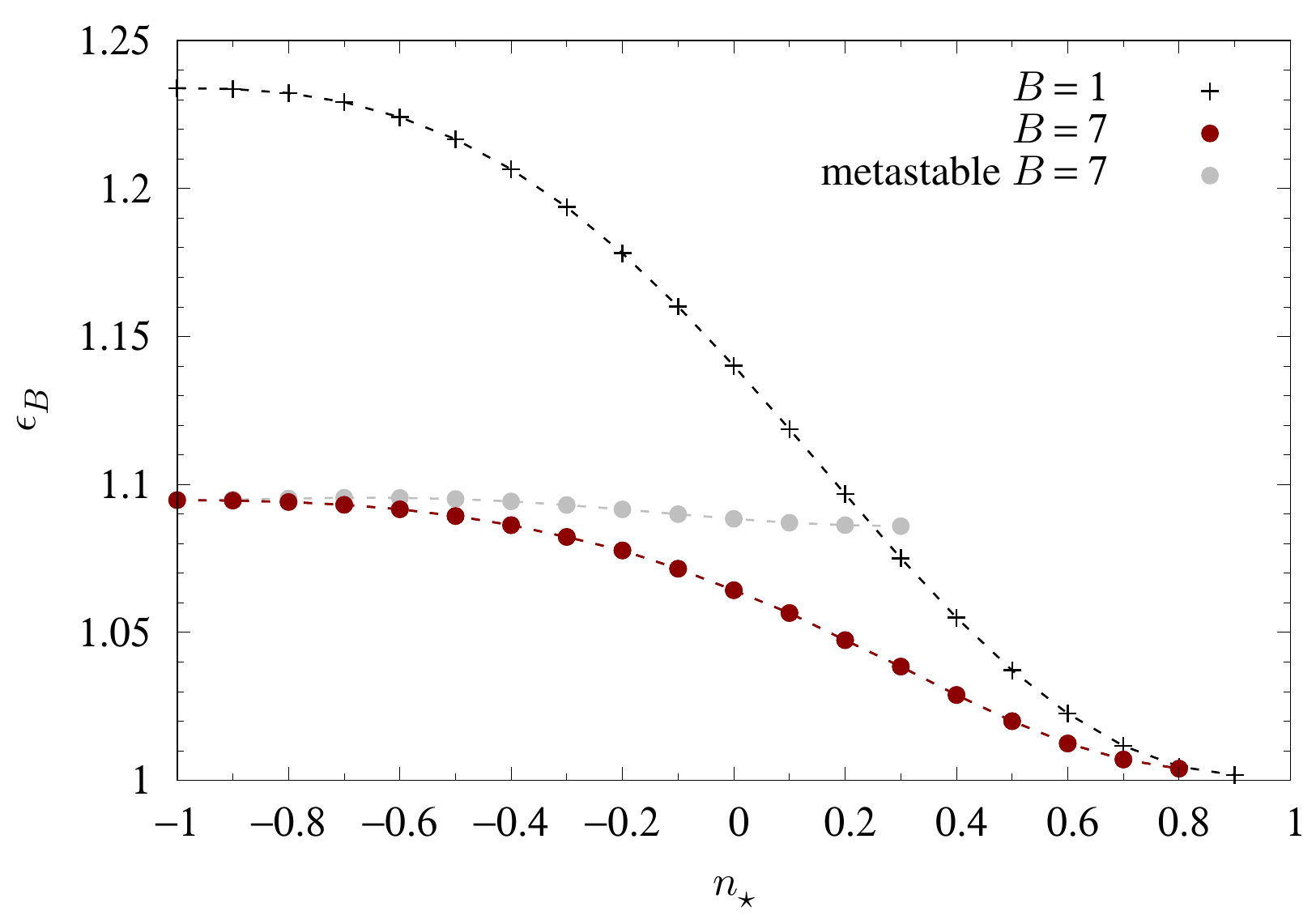}}}
    \caption{Energy of all \emph{massless} Skyrmion solutions,
      including metastable states (gray lines), normalized by their
      respective Bogomol'nyi bounds, $\epsilon_B$ of
      eq.~\eqref{eq:epsilon_B}: for baryon number (a) $B=2$ to (f)
      $B=7$.
      For comparison, the normalized energies of the $B=1$ Skyrmion
      are shown as well.
    }
    \label{fig:en2-7}
  \end{center}
\end{figure}

Fig.~\ref{fig:en2-7} shows the energies, topological sector by
topological sector, comparing the normalized energy $\epsilon_B$ of
the $B$-Skyrmion with the 1-Skyrmion, for $B=2$ through $B=7$.
With the exception of $B=2$, all the $B$-Skyrmions possess a branch of
metastable solutions which look exactly like the $n_\star=-1$ solution
(up to a slight change in length scale).
We might say that the higher the discrete symmetry is, the longer up
in $n_\star$ the metastable branch of solutions exists; see $B=4,6,7$
in fig.~\ref{fig:en2-7}(c),(e),(f).
The metastable branch for $B=7$ is particularly long lived, so long
that it even crosses the energy curve of the 1-Skyrmion: this means
that energy could be gained at that point, by splitting the solution
up to seven well-separated 1-Skyrmions.
This energy, as we will discuss shortly, is called (minus) the binding
energy.

\begin{figure}[!htp]
  \begin{center}
    \mbox{\subfloat[]{\includegraphics[width=0.49\linewidth]{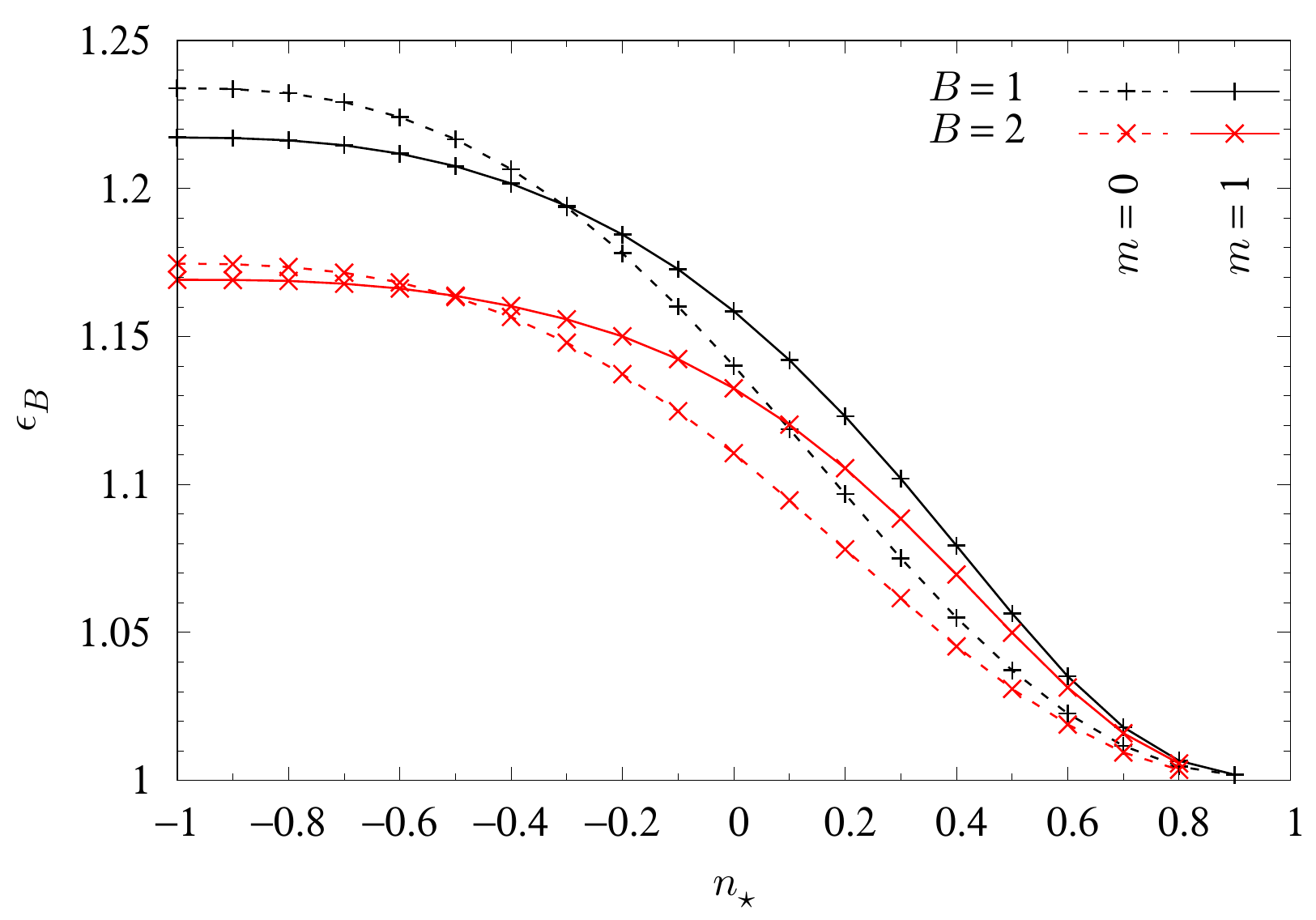}}
      \subfloat[]{\includegraphics[width=0.49\linewidth]{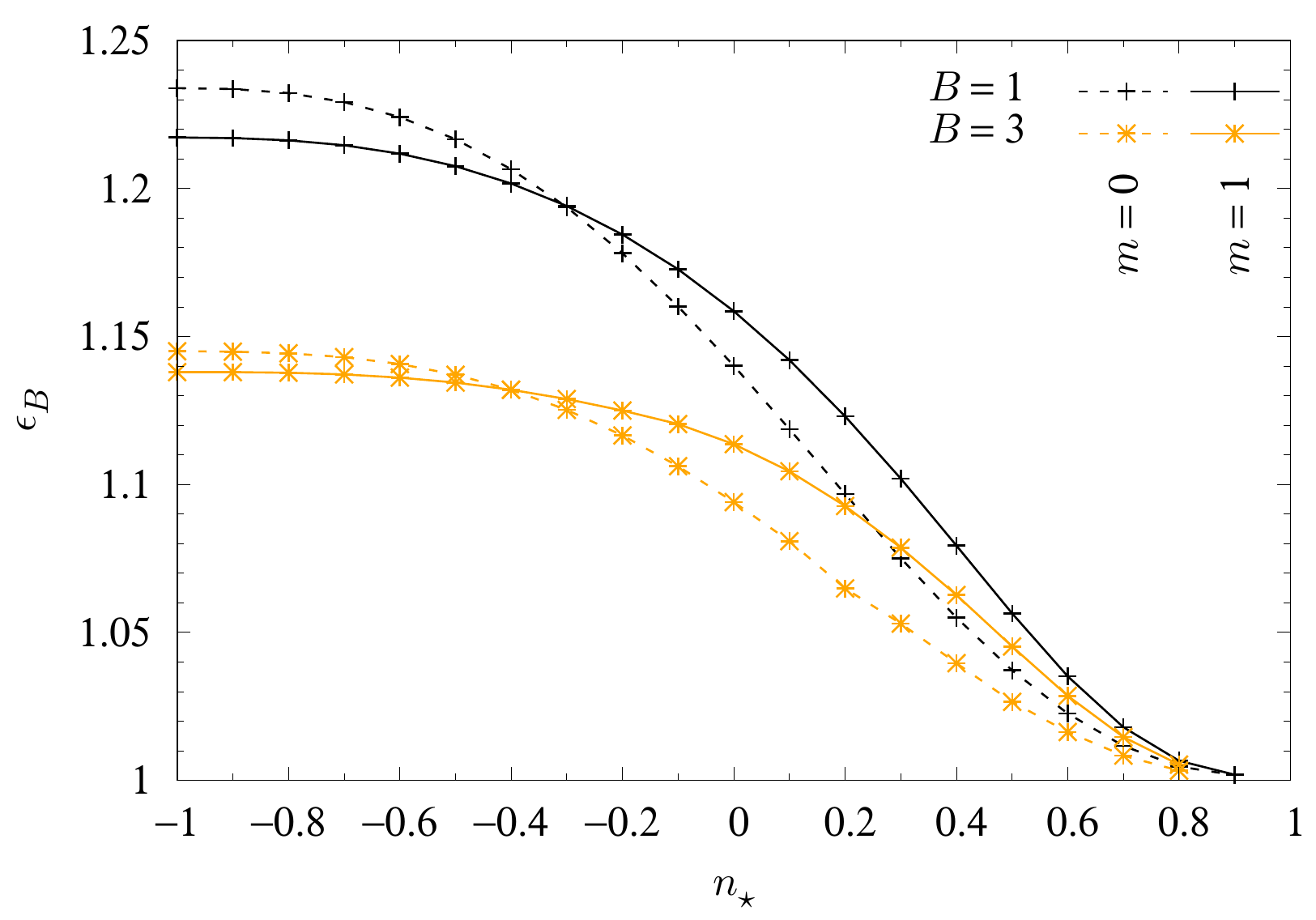}}}
    \mbox{\subfloat[]{\includegraphics[width=0.49\linewidth]{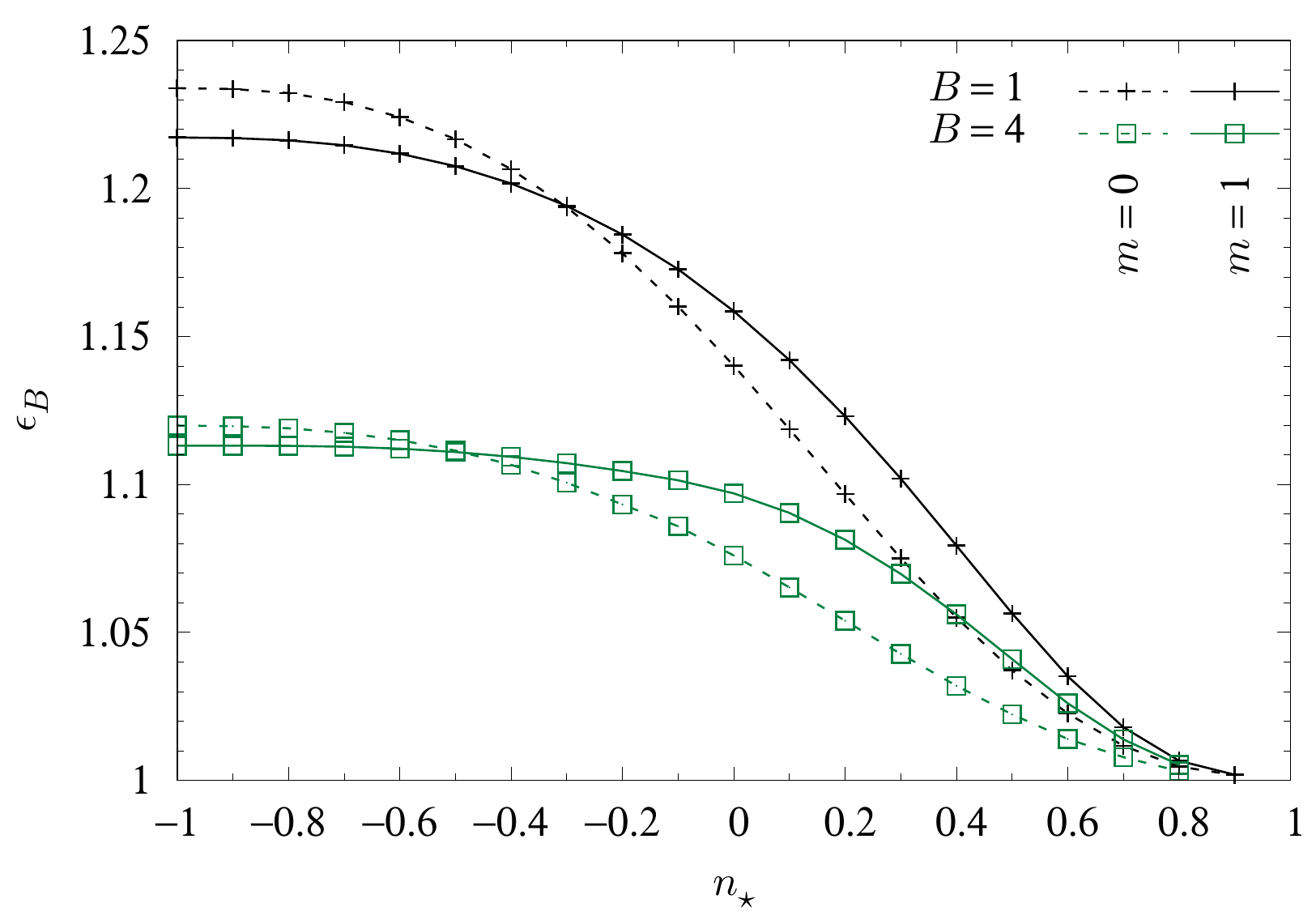}}
      \subfloat[]{\includegraphics[width=0.49\linewidth]{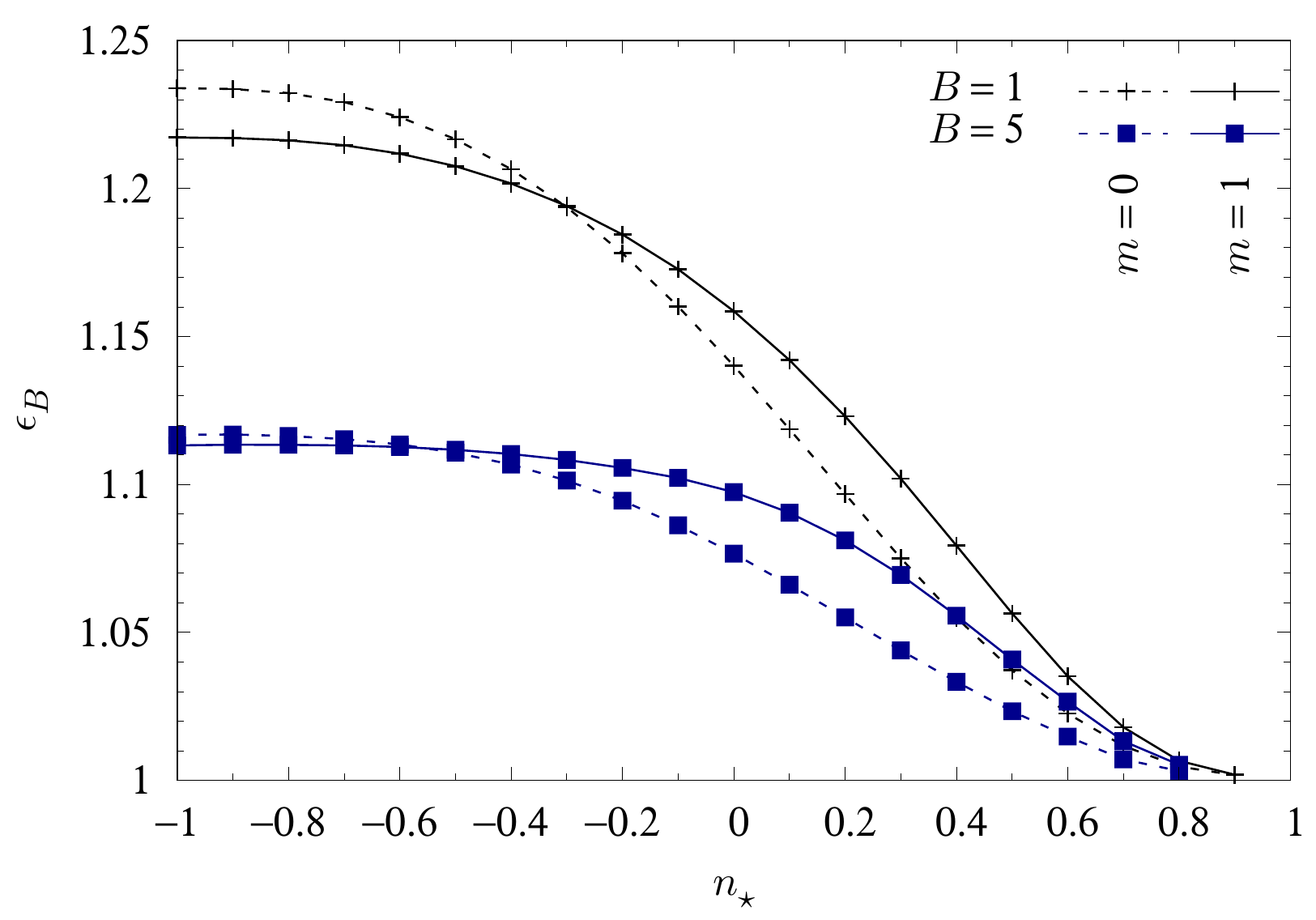}}}
    \mbox{\subfloat[]{\includegraphics[width=0.49\linewidth]{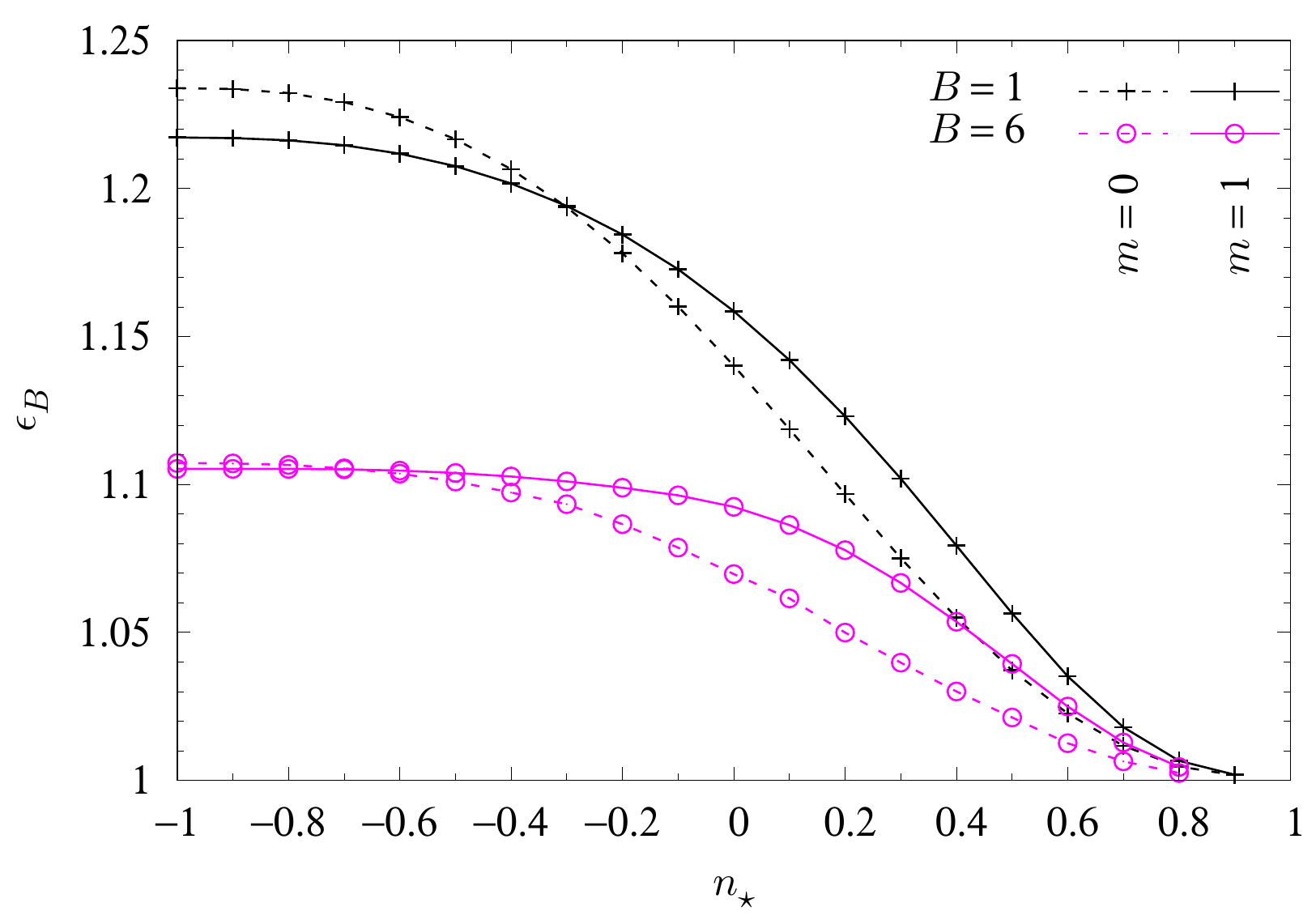}}
      \subfloat[]{\includegraphics[width=0.49\linewidth]{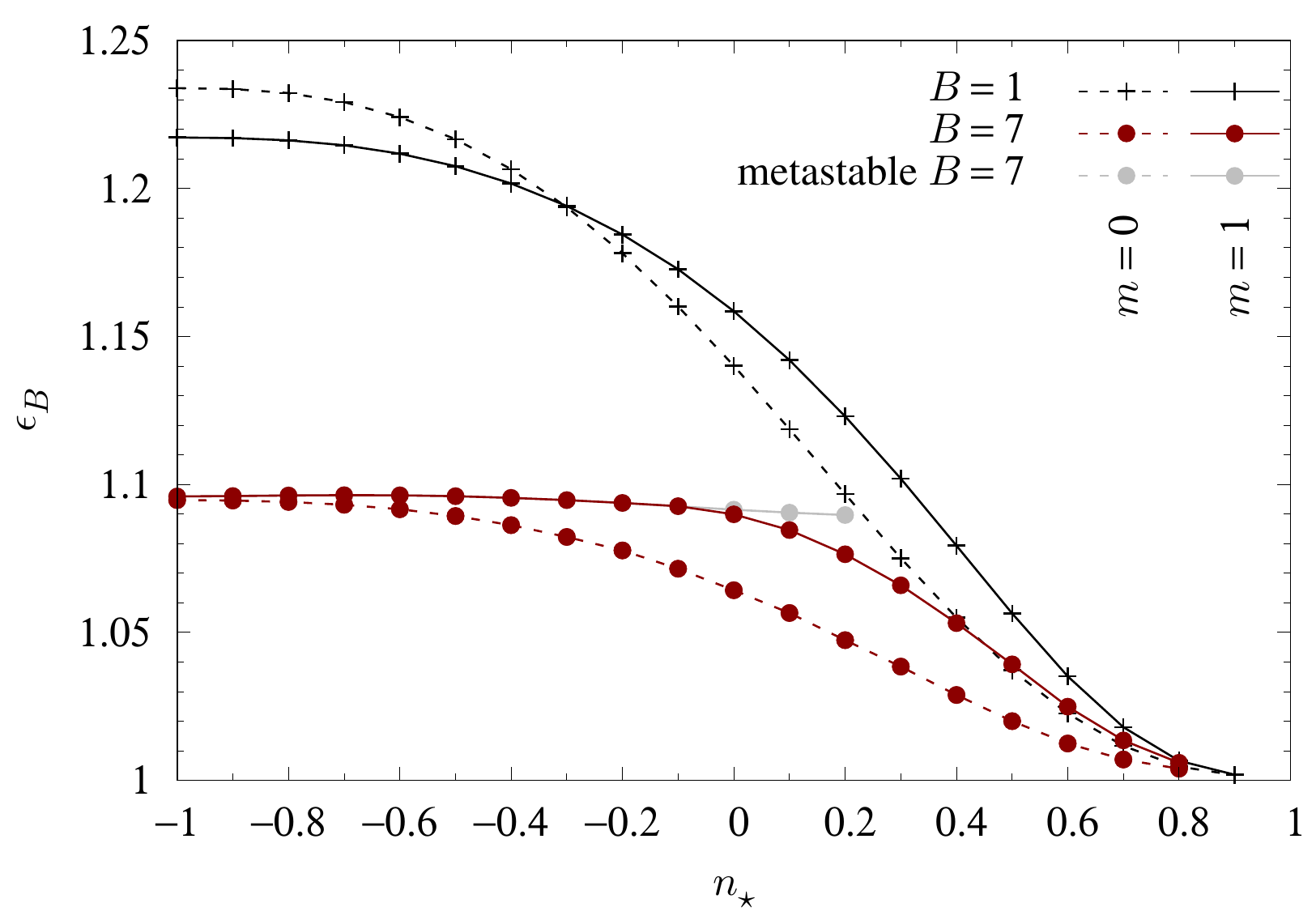}}}
    \caption{Energy of all \emph{massive} Skyrmion solutions,
      including metastable states (gray lines), normalized by their
      respective Bogomol'nyi bounds, $\epsilon_B$ of
      eq.~\eqref{eq:epsilon_B_m1}: 
      for baryon number (a) $B=2$ to (f) $B=7$.
      For comparison, the normalized energies of the massless
      $B$-Skyrmions and the massive 1-Skyrmion are shown as well.
    }
    \label{fig:en2-7m1}
  \end{center}
\end{figure}

We will now turn to the \emph{massive} case ($m=1$), for which the
normalized energies \eqref{eq:epsilon_B_m1} are shown in
fig.~\ref{fig:en2-7m1} with solid lines, for topological charges $B=2$
through $B=7$.
For comparison with the massless case ($m=0$), we have kept the
energies of the \emph{stable} solutions in each figure, shown with
dashed lines.
In every figure, the massive and the massless solutions are also
compared with the $B=1$ energies.
The difference on the $y$-axis between the normalized $B$-Skyrmion
energy and the normalized 1-Skyrmion energy, is exactly the binding
energy per baryon (nucleon).
Hence, as long as the normalized $B$-Skyrmion energy is below the
normalized 1-Skyrmion energy, the $B$-Skyrmion is bound and thus at
least cannot break up into 1-Skyrmions without an additional kick
(external energy).
Thus, all the solutions in the entire fig.~\ref{fig:en2-7m1} are
bound.
In fig.~\ref{fig:en2-7m1}, we also show the metastable solutions with
gray solid lines, but in contrast to the massless case, there are no
metastable branches, except for the $B=7$ solution, see
fig.~\ref{fig:en2-7m1}(f). 

A comment in store is about the absolute (unnormalized) energies of
the massive Skyrmion solutions.
That is, although it seems that all the massive solutions have smaller
energies at $n_\star=-1$ than the corresponding massless solutions;
that is an artifact of the different normalizations: the massless
energies are normalized according to eq.~\eqref{eq:epsilon_B}, whereas
the massive energies are normalized according to
eq.~\eqref{eq:epsilon_B_m1}.
In reality, the massive Skyrmion solutions are always heavier in
Skyrme units than the corresponding massless solutions, but it turns
out that for $B<7$ they are closer to their respective energy bound
than the massless ones.

\begin{figure}[!htp]
  \begin{center}
    \mbox{\subfloat[$m=0$]{\includegraphics[width=0.49\linewidth]{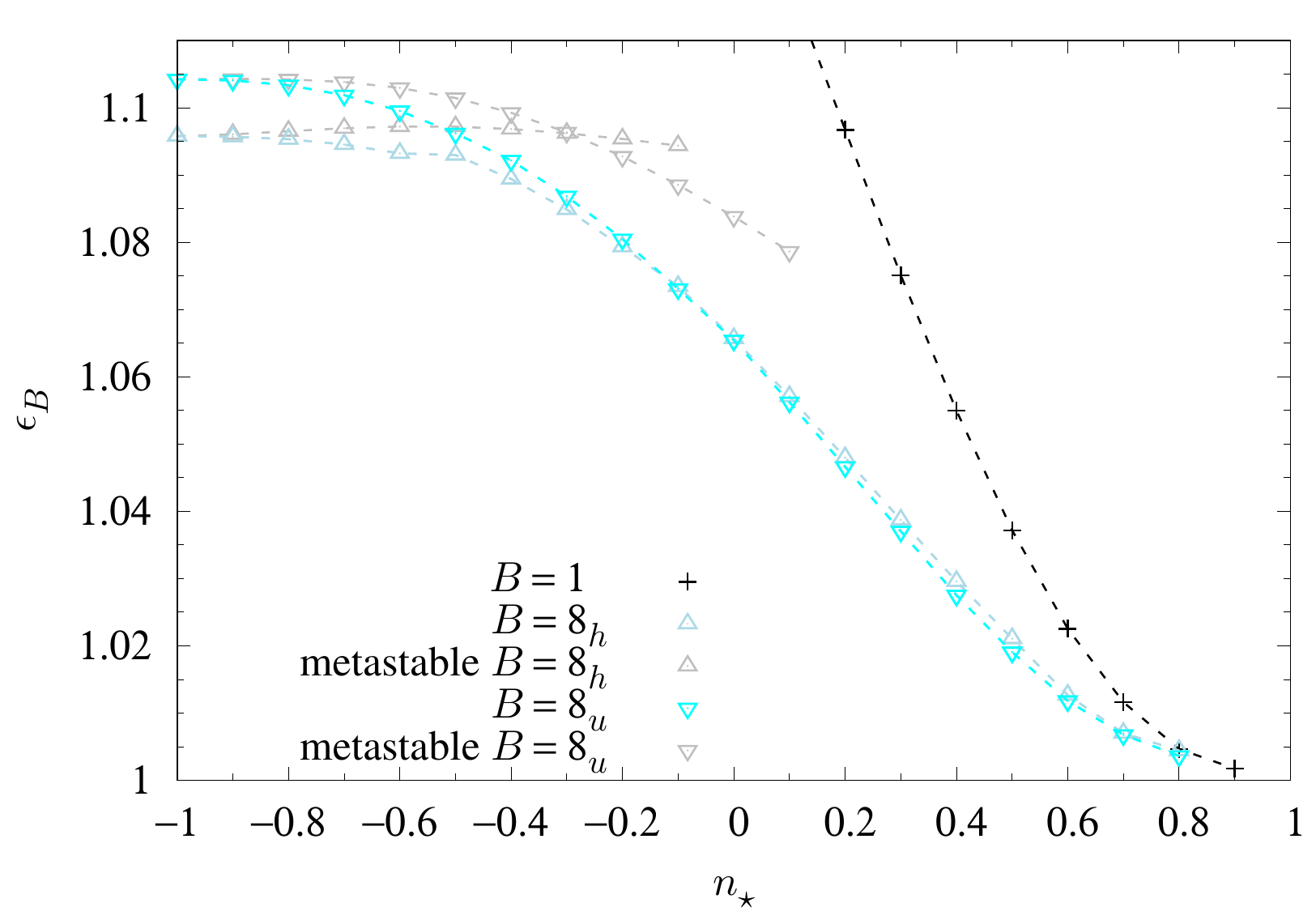}}
      \subfloat[$m=1$]{\includegraphics[width=0.49\linewidth]{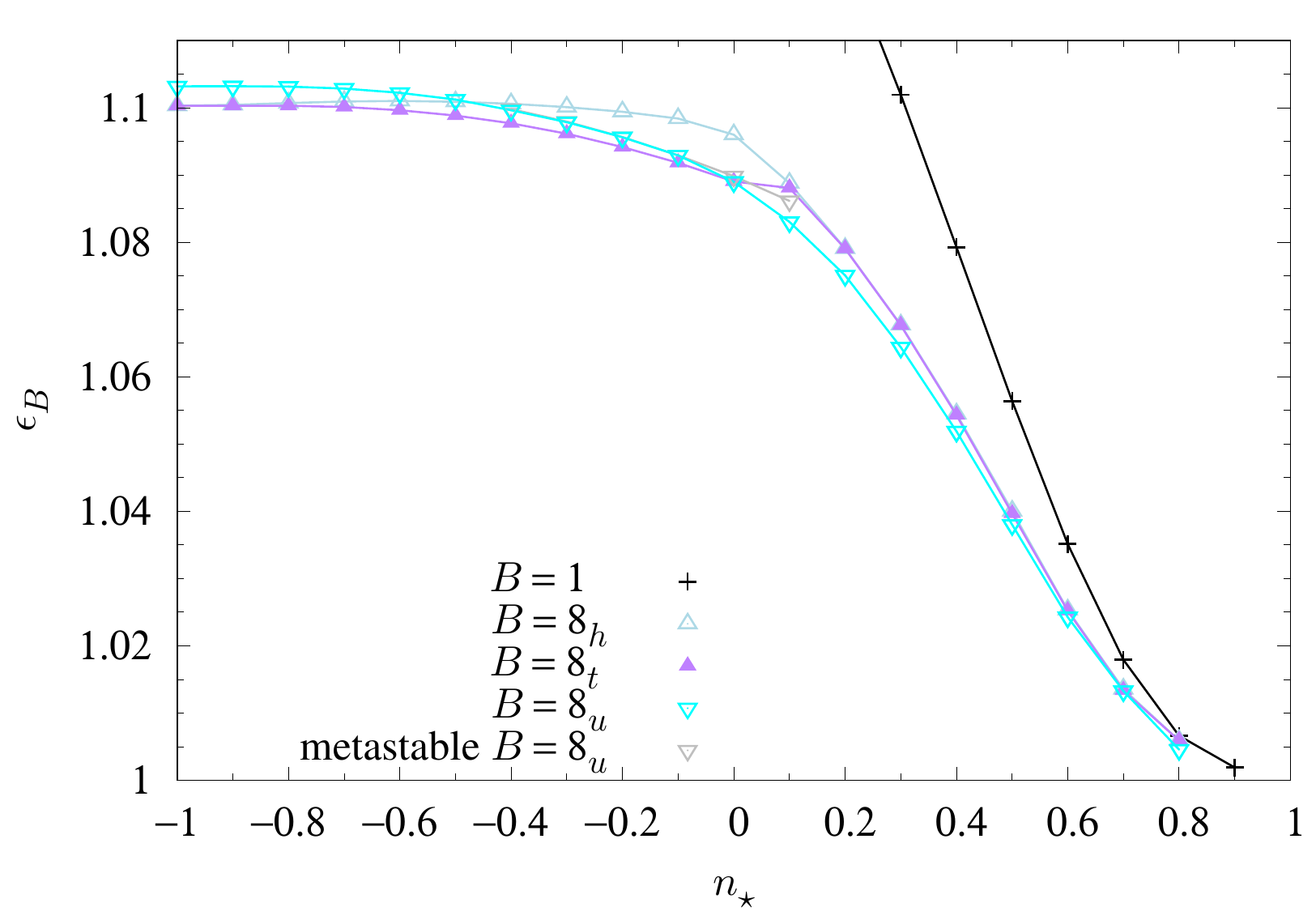}}}
    \caption{Energy of all Skyrmion solutions normalized by their
      respective Bogomol'nyi bounds, $\epsilon_B$ for $B=8$: (a)
      eq.~\eqref{eq:epsilon_B}: without the pion mass term, (b)
      eq.~\eqref{eq:epsilon_B_m1}: with the pion mass term and $m=1$. 
    }
    \label{fig:en8}
  \end{center}
\end{figure}

We now come to the $B=8$ sector, which is the most complicated
topological sector that we study in this paper, simply because there
are more than one classical (metastable) solution.
Fig.~\ref{fig:en8} shows the $B=8$ sector for both the massless case
(fig.~\ref{fig:en8}(a)) and the massive case (fig.~\ref{fig:en8}(b)).
Although we have kept the normalized energies of the $B=1$ Skyrmion,
we have zoomed the figure so as to better display the features of the
$B=8$ energies.

Starting with the massless case $(m=0)$ in fig.~\ref{fig:en8}(a), we
have two solutions with different symmetries and hence different
shapes, which we shall refer to as the $B=8_h$ for the dihedrally
symmetric Skyrmion depicted in fig.~\ref{fig:mon6-8u}(c) as well as
the $B=8_u$ denoting the untwisted chain with cubic symmetry, depicted
in fig.~\ref{fig:mon6-8u}(d).
Both of them have metastable branches of solutions, where they retain
the discrete symmetry present for $n_\star=-1$, which are dihedral and
cubic symmetry for the $B=8_h$ and $B=8_u$, respectively.
For $n_\star=-1$, which is the standard massless Skyrme model, the
stable Skyrmion in the $B=8$ sector is the $B=8_h$ dihedrally
symmetric one, with $\epsilon_B=1.0958$ compared to
$\epsilon_B=1.1043$ for the $B=8_u$ -- and so ($B=8_h$) is very
clearly the stable solution.
An interesting twist happens when increasing $n_\star$, which is that
the $B=8_u$ untwisted chain Skyrmion becomes the stable solution
for large $n_\star$ and the crossover point is around
$n_\star\sim -0.1$.
For $n_\star\gtrsim 0.6$, the difference in energy between the two
solutions becomes quite small though.

Turning now to the massive case ($m=1$), we have three solutions: the
dihedrally symmetric $B=8_h$, the twisted chain $B=8_t$, with cubic
symmetry and the untwisted chain $B=8_u$, also with cubic symmetry,
see figs.~\ref{fig:mon6-8u}(c), \ref{fig:mon8tm1}, and
\ref{fig:mon6-8u}(d). 
The stable solution is probably the twisted chain $B=8_t$, in accord
with the findings of ref.~\cite{Battye:2006na}, but as stated in the
latter reference, the difference in energy between the twisted chain
($B=8_t$) and the dihedrally symmetric ($B=8_h$) Skyrmion is so small
(for $m=1$) that it is beyond the numerical accuracy to decide which
one is truly the global minimizer of the energy -- and it would be
consistent with our results if they are degenerate.
We find that $\epsilon_B=1.10030$ for the $B=8_t$ Skyrmion, compared
to $\epsilon_B=1.10031$ for the $B=8_h$ Skyrmion\footnote{Notice that
  ref.~\cite{Battye:2006na} used the energy bound \eqref{eq:epsilon_B}
for the massless Skyrme model (at $n_\star=-1$), whereas we are
normalizing the energy by the bound \eqref{eq:epsilon_B_m1} for the
\emph{massive} Skyrme model (at $n_\star=-1$). }.
The almost degenerate-in-energy solutions $B=8_t$ and $B=8_h$ are
clearly lower in energy with respect to the $B=8_u$ solution, see
fig.~\ref{fig:en8}(b).

Upon increasing $n_\star$ from $n_\star=-1$, the twisted chain
($B=8_t$) remains the stable solution (and more convincingly so),
until $n_\star=0$, where the untwisted chain crosses over and becomes
the stable solution throughout the range up to $n_\star=0.8$ (the
maximum of $n_\star$ in our calculations).
Around the point $n_\star=0$, what happens is that the 1-Skyrmions
become localized and the model is turned into a point-particle model
of Skyrmions.
Interestingly, the dihedrally symmetric $B=8_h$ solution, initially
increases in energy with $n_\star$ increasing from $n_\star=-1$, then
crosses over the previously highest-in-energy $B=8_u$ solution at
$n_\star\sim-0.5$ to become the least stable solution.
However, from $n_\star\geq 0.1$, the $B=8_h$ and $B=8_t$ solution
fall into an identical pattern of point-particle Skyrmions sitting at
the vertices of an FCC lattice, which is reminiscent of the dihedrally 
symmetric Skyrmion solution.
The untwisted chain, $B=8_u$, on the other hand, upon dissolving into
point-particle Skyrmions, falls into a pattern of two weakly
interacting tetrahedrally arranged $B=4$ Skyrmions, which is a
different pattern of eight point particles placed at the vertices of
an FCC lattice, but with lower energy.

\begin{figure}[!htp]
  \begin{center}
    \mbox{\subfloat[$m=0$]{\includegraphics[width=0.49\linewidth]{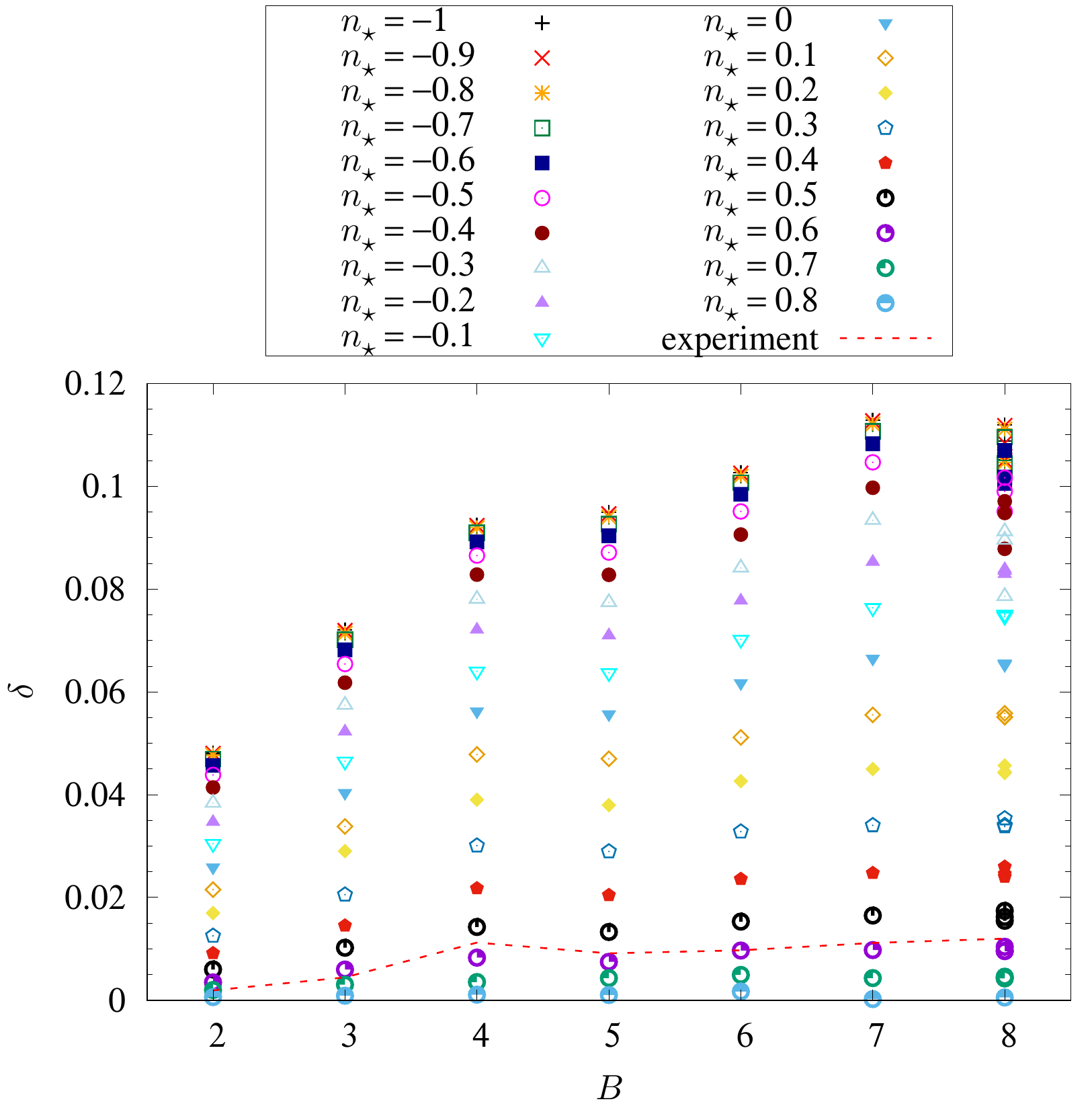}}
      \subfloat[$m=1$]{\includegraphics[width=0.49\linewidth]{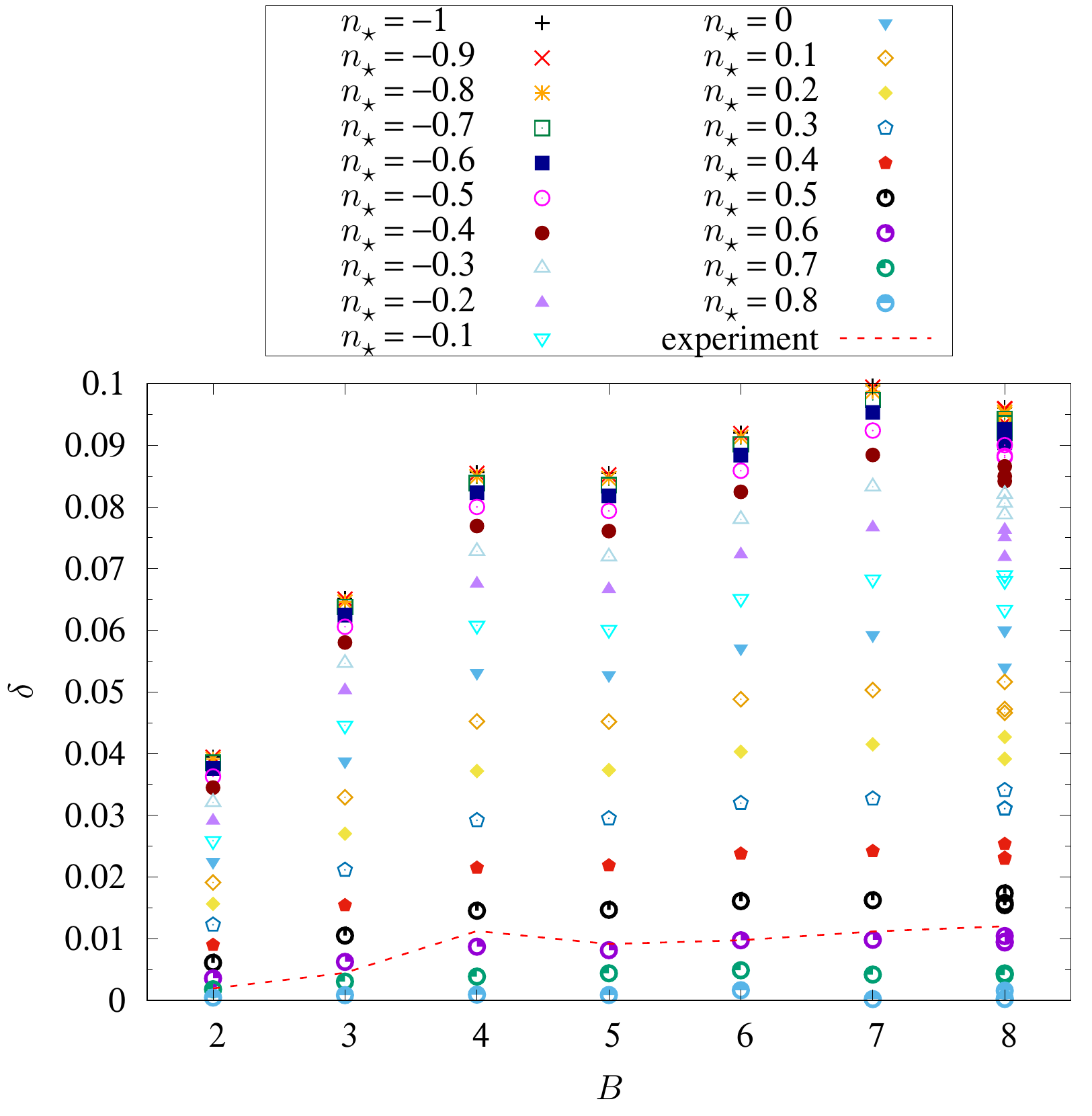}}}
    \mbox{\subfloat[$m=0$]{\includegraphics[width=0.49\linewidth]{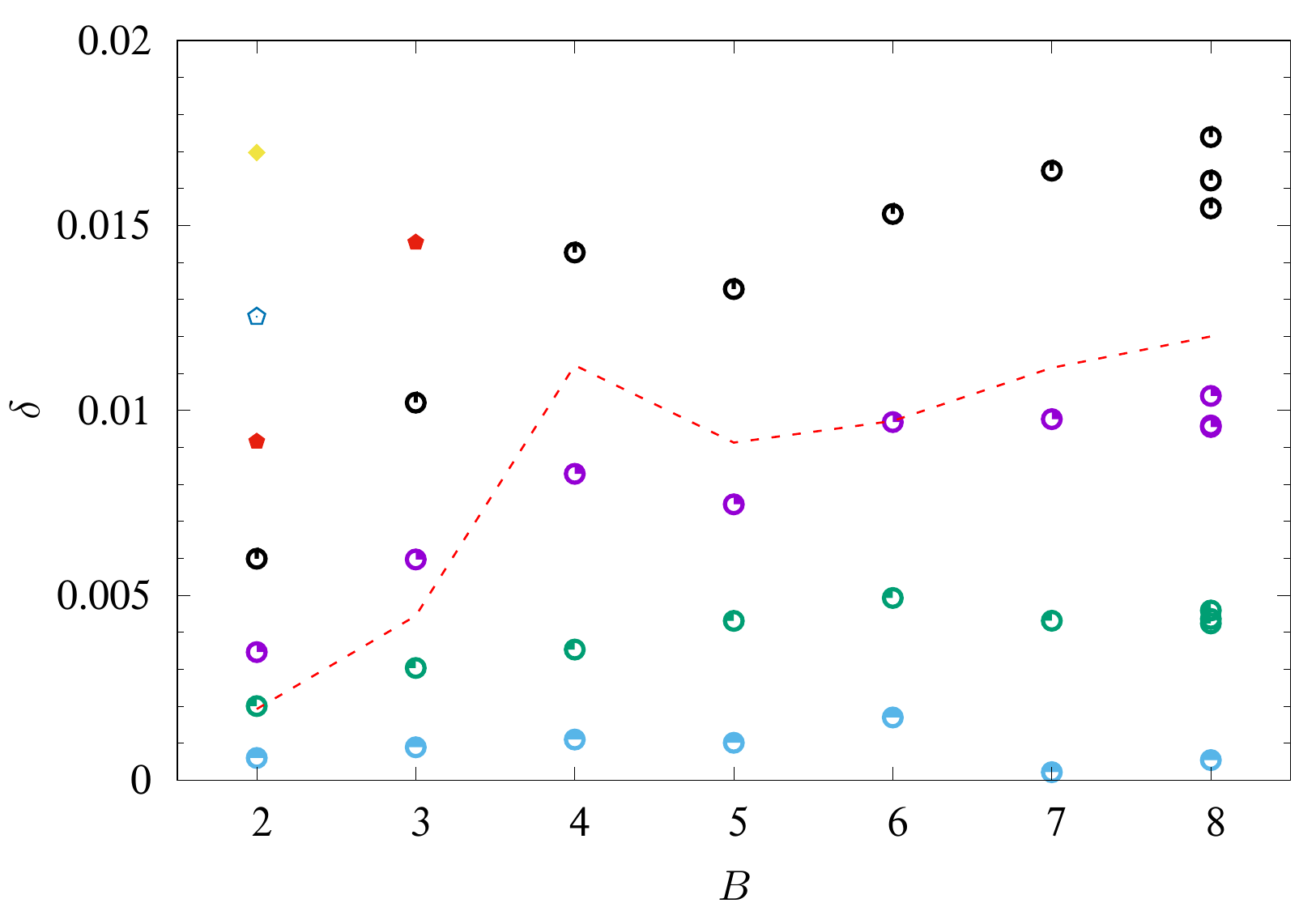}}
      \subfloat[$m=1$]{\includegraphics[width=0.49\linewidth]{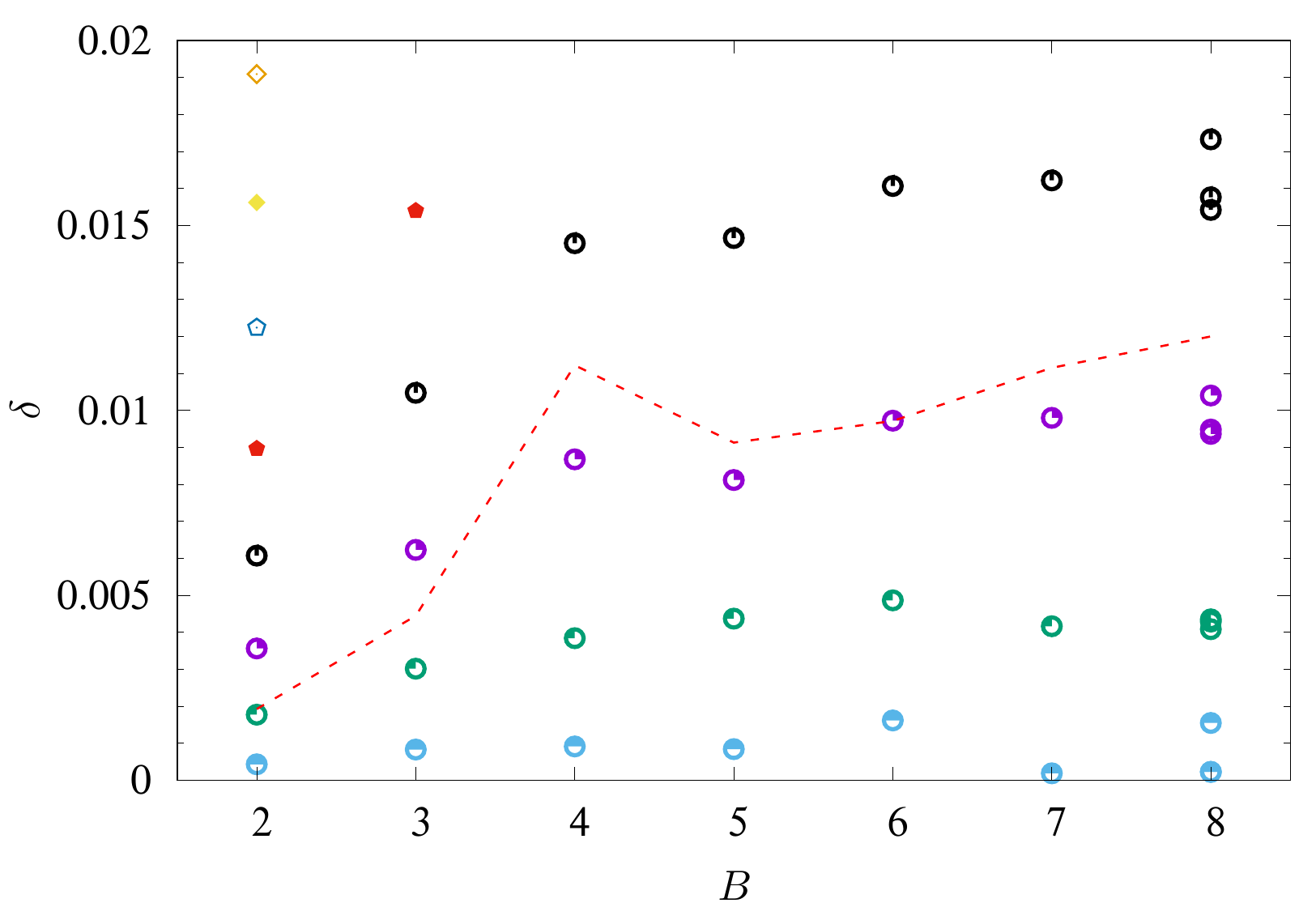}}}
    \caption{Relative binding energies, $\delta$, of all Skyrmion
      solutions (a) in the massless case ($m=0$) and (b) in the
      massive case ($m=1$).
      The panels (c) and (d) display zoom-ins of the low-binding
      energy part of the panels (a) and (b), respectively. 
    }
    \label{fig:ben}
  \end{center}
\end{figure}

We are now finally ready to discuss the relative binding energies,
defined by
\beq
\delta = \frac{BE_1 - E_B}{BE_1}
= 1 - \frac{E_B}{BE_1},
\label{eq:delta}
\eeq
which is a dimensionless quantity and independent of units or
normalization of the energies in question.

The relative binding energies, $\delta$ of eq.~\eqref{eq:delta}, are
displayed in fig.~\ref{fig:ben} for all the Skyrmion solutions in the
near-BPS dielectric Skyrme model.
The massless solutions are shown in fig.~\ref{fig:ben}(a) and the
massive solutions in fig.~\ref{fig:ben}(b).
For the massless case, $\delta$ ranges from about 11.5\% down very
close to zero, whereas in the massive case it ranges from about 10\%
again down very close to zero.
For small values of $n_\star$ -- which correspond to the almost
standard Skyrme model, the inclusion of the pion mass thus decreases
the binding energies a bit, as known in the literature.
For higher values of $n_\star$, the binding energies drop quite
dramatically and for that reason, we have added two extra panels to
the figure, zooming in on the low-binding energy region in
figs.~\ref{fig:ben}(c) and \ref{fig:ben}(d). 
As a guide we have added a red-dashed line depicting the
experimentally observed binding energies for nuclei on all panels in
fig.~\ref{fig:ben}.
We can see that the closest set of Skyrmion solutions is that
corresponding to $n_\star=0.6$ and some features of physics can be
qualitatively seen from the $n_\star=0.6$ data points: For example,
the relative binding energy increases for $B=2,3,4$ culminating in the
stable $\alpha$ particle, but drops for $B=5$ -- which in Nature is
unstable.
We will discuss the results further in the next section.

\section{Discussion}\label{sec:discussion}

In this paper we have considered a near-BPS version suggested in
ref.~\cite{Adam:2020iye} of the dielectric Skyrme model, which is yet
another modification -- albeit a bit drastic one at that -- of the
standard Skyrme model, in order to lower the binding energies. 
After reviewing the model and re-deriving the Bogomol'nyi bound, we
specify the choice of BPS breaking in terms of the function $f(n_0)$.
We further include the case of adding the pion mass term to the
Lagrangian -- although that does not logically guarantee a limit with
small binding energies.
Finally, we perform a numerical investigation of the Skyrmion
solutions in the topological charge sectors $B=1$ through $B=8$, with
the initial conditions taken from the rational map approximations and
for $B=8$ two additional initial conditions made of $B=4$ cubes.
First of all we find that, the near-BPS limit studied in this paper,
reduces the solutions to the point-particle Skyrmions obtained in
other models in the literature.
Finally, we demonstrate that realistic classical binding energies can
be obtained in this near-BPS model too. 

For obtaining a completely realistic theoretical curve for the
relative binding energies, further effects must be taken into account,
of course.
Crucially, of course, is the quantization of spin and isospin, which is a
necessity for talking about fermionic states at all.
Furthermore, Coulomb energy and isospin breaking effects must be
considered too, if fine details of the curve should be trusted.
Such refinements are beyond the scope of this paper though.
The punchline is that we have lowered the classical binding energies
to below the experimentally measured ones and the remaining problem for
using the Skyrme model as a model for nuclear theory lies entirely in
developing a suitable quantization scheme.

We would like to comment that the commonly known problems with rigid-body
quantization might not be as severe in the point-particle models of
Skyrmions.
The reason is simply that quantizing the cluster of localized
1-Skyrmions as a rigid rotor physically makes no sense.
If the overlaps of the tails of the 1-Skyrmions are sufficiently
small, it is conceivable that the quantum correction to the energy due
to spin and isospin is closer to $B$ times that of the 1-Skyrmions, as
compared to the problem in the standard Skyrme model, where the
correction is large for the 1-Skyrmion and almost vanishing for $B>1$.
This claim needs further investigation and should be part of the
development of a mature quantization scheme. 

One might naively think that there are many more effects from nuclear
physics that must be incorporated into the Skyrme model, in order to
produce a working theoretical precision framework for nuclear theory.
For example, 3-body forces are known to be important in conventional
nuclear theory for lowering binding energies and among many
applications, reproducing the oxygen drip line, see
e.g.~ref.~\cite{Otsuka:2010}.
An interpretation of this is that a repulsive contribution to the
binding energy is coming from the two-pion exchange captured by the
3-body force \cite{Otsuka:2010}.
In the Skyrme-type models, however, all-body forces are in principle
taken into account and such additions of particular effects are not
necessary, but are supposed to be already built-in.
Other forces, like the spin-orbit force, have also recently been shown
to be already accounted for correctly in the Skyrme model
\cite{Halcrow:2020gbm} -- in contradistinction with early findings. 

Since many Skyrme-type models have now been obtained, successfully
lowering the classical binding energies to realistic levels, see 
refs.~\cite{Gillard:2015eia,Gillard:2016esy,Gudnason:2018jia,Gudnason:2020arj}
(including the model in the present paper), the time is perhaps ripe
to turn to the development of a suitable quantization scheme for
applying (some version of) Skyrme theory to nuclear physics.

\subsection*{Acknowledgments}

S.~B.~G.~thanks the organizers of
``Various topics in mathematical physics,'' for the stimulating online
virtual conference environment and Andrzej Wereszczynski for giving a
talk that motivated this work. 
S.~B.~G.~thanks the Outstanding Talent Program of Henan University for
partial support.
The work of S.~B.~G.~is supported by the National Natural Science
Foundation of China (Grants No.~11675223 and No.~12071111).

\end{document}